\documentclass[twocolumn]{aastex701}
\usepackage{amsmath}
\usepackage{siunitx} 

\usepackage{booktabs}
\usepackage{longtable}
\usepackage{threeparttable}
\usepackage{array}
\usepackage{multirow}
\usepackage{CJK}
\usepackage{lineno}
\newcommand{\swiftbat}[1]{\textit{Swift}-BAT}
\newcommand{\swiftxrt}[1]{\textit{Swift}-XRT}
\newcommand{\batxrt}[1]{\textit{Swift}-BAT/XRT}
\newcommand{\swiftUVOT}[1]{\textit{Swift}-UVOT}
\newcommand{\fermigbm}[1]{\textit{Fermi}-GBM}
\newcommand{\fermilat}[1]{\textit{Fermi}-LAT}
\linenumbers
\begin{document}

\title{Multi-Messenger Modeling of Low-Luminosity $\gamma$-Ray Bursts}

\author[orcid=0000-0003-0035-7766]{Shiqi Yu \begin{CJK*}{UTF8}{gbsn}(于世琦)\end{CJK*}}
\affiliation{Department of Physics and Astronomy, University of Utah, Salt Lake City, Utah, USA, 84102}
\email[show]{shiqi.yu@utah.edu}  

\author[orcid=0000-0003-2478-333X]{B. Theodore Zhang \begin{CJK*}{UTF8}{gbsn}(张兵)\end{CJK*}}
\affiliation{Key Laboratory of Particle Astrophysics and Experimental Physics Division and Computing Center, Institute of High Energy Physics, Chinese Academy of Sciences, 100049 Beijing, China}
\affiliation{TIANFU Cosmic Ray Research Center, Chengdu, Sichuan, China}
\email[show]{zhangbing@ihep.ac.cn}  
\begin{abstract}

Low-luminosity gamma-ray bursts (LL GRBs), a subclass of the most powerful transients in the Universe, remain promising sources of high-energy astrophysical neutrinos, despite strong IceCube constraints on typical long GRBs. In this work, a novel approach is introduced to study a sample of seven LL~GRBs with their multi-wavelength observations to investigate leptohadronic processes during their prompt emission phases. 
The relative energy densities in magnetic fields, non-thermal electrons, and protons are constrained, with the latter defining the cosmic-ray (CR) loading factor. Our results suggest that LL~GRBs exhibit diverse emission processes, as confirmed by a machine-learning analysis of the fitted parameters. Across the seven LL~GRBs, we find the posterior medians of the CR loading factor in the range of $\xi_p \sim 0.2$--$1.6$. GRB~060218 and GRB~100316D, the lowest-luminosity bursts ($L_{\gamma, \rm iso} \sim 10^{46}$-$10^{47}\rm~erg~s^{-1}$) consistent with the shock-breakout (SBO) scenario, yield the highest CR loading factor and therefore are expected to produce neutrinos more efficiently. Our model predicts the expected number of neutrino signals that are consistent with current limits but would be detectable with next-generation neutrino observatories. These results strengthen the case for LL~GRBs as promising sources of high-energy astrophysical neutrinos and motivate real-time searches for coincident LL~GRB and neutrino events. Next-generation X-ray and MeV facilities will be critical for identifying more LL~GRBs and strengthening their role in multi-messenger astrophysics.

\end{abstract}


\section{Introduction} \label{sec:intro}
The diffuse flux of high-energy astrophysical neutrinos in the TeV-PeV range has been reported by multiple studies~\citep{ic_diffuse_2013,PRL125-121104,Aartsen_2015,IceCube:2020wum,Abbasi_2022,IceCube:2024fxo,bpl_icecube25,IceCube:2025tgp}, yet their astrophysical origin remains largely unknown. While some of the active galactic nuclei (AGN) have been identified as sources of high-energy neutrinos, including the Seyfert galaxy NGC~1068~\citep{sci2022_icecube} and the blazar TXS~0506+056~\citep{IceCube:2018cha}, several studies indicate that they cannot account for the full observed diffuse flux, particularly toward $\gtrsim$0.1 PeV energies~\citep[e.g., ][]{Kimura:2014jba, Fang:2017zjf, Murase:2019vdl, McDonough:2023ngk, Saurenhaus:2025ysu}. This motivates the exploration of additional source classes, such as low-luminosity gamma-ray bursts (LL~GRBs).

GRBs have long been considered promising sources of multiwavelength electromagnetic radiation~\citep{Zhang_2009}, cosmic rays~\citep{Waxman:1995vg, Vietri_1995}, and high-energy neutrinos~\citep{Waxman:1997ti}. Several theoretical models have been proposed to explain the prompt emission, including internal shocks~\citep{Rees:1992ek}, photospheric emission~\citep{Rees:2004gt}, and the Internal-Collision-Induced Magnetic Reconnection and Turbulence (ICMART) model~\citep{Zhang:2010jt}. In the prompt emission, accelerated protons can interact with ambient photons to produce charged pions and kaons, which decay into high-energy neutrinos~\citep[e.g.,][]{Waxman:1997ti}. The diffuse neutrino flux from GRBs is estimated to be $E_\nu^2 \phi_\nu \sim 10^{-8}\rm~GeV~cm^{-2}~s^{-1}~sr^{-1}$ at $E_\nu\sim$ 100 TeV~\citep{Waxman_1998}, comparable to the flux later observed by IceCube~\citep{IceCube:2020wum,Abbasi_2022}.

LL~GRBs are roughly 100 times less energetic than typical high-luminosity (HL) GRBs, with $L_{\gamma, \rm iso} \lesssim 10^{49}\rm~erg~s^{-1}$ or $E_{\gamma, \rm iso} \sim 10^{48}-10^{50}\rm~erg$, and feature trans-relativistic ejecta. They are 10-1000 times more abundant than HL GRBs and exhibit longer durations~\citep{Soderberg_2006, Liang:2006ci}, making them promising contributors to the diffuse high-energy neutrino flux. Because only a limited number of LL~GRBs have been detected, their event rate and physical properties are highly uncertain~\citep{Sun:2015bda}, and current observational estimates may underestimate their contribution.

Searches for neutrino emission from the prompt phase of long GRBs using the fireball model have placed stringent constraints, limiting the contribution of typical HL GRBs to at most $\sim 1\%$ of the observed diffuse flux~\citep{Aartsen_2017, IceCube:2012qza}. While some LL GRBs were included in the long GRB sample, the constraints are dominated by the HL GRBs. Nevertheless, LL~GRBs may still contribute significantly to the observed diffuse high-energy neutrino flux.

In this work, we perform a comprehensive study of photon and neutrino emission from LL~GRBs. We analyze a sample of observed LL~GRB candidates, discuss their physical origins in Sec.~\ref{sec:cat}, and present the numerical methods and leptohadronic modeling in Sec.~\ref{sec:model}. Multiwavelength observations are analyzed in Sec.~\ref{sec:mm} to constrain the physical parameters relevant to hadronic processes. Implications are discussed in Sec.~\ref{sec:result_discussion}, and Sec.~\ref{sec:summary} summarizes our findings and future perspectives.

Throughout this study, we adopt cgs (centimeter-gram-second) units and express physical quantities in the form \( Q_x = Q \times 10^x \). Particle energy is denoted by \( E \) in the observer frame and by \( \varepsilon^\prime \) in the comoving frame. 
We adopt a $\Lambda$CDM cosmology with $H_0 = 71\rm~km~s^{-1}~Mpc^{-1}$, $\Omega_M  = 0.27$, and $\Omega_\Lambda = 0.73$~\cite{ParticleDataGroup:2024cfk}.

\section{Low-luminosity GRBs}\label{sec:cat}

\begin{figure*}[!tbh]
    \centering 
\includegraphics[width=0.32\linewidth,trim=40cm 19cm 50cm 10cm,clip]{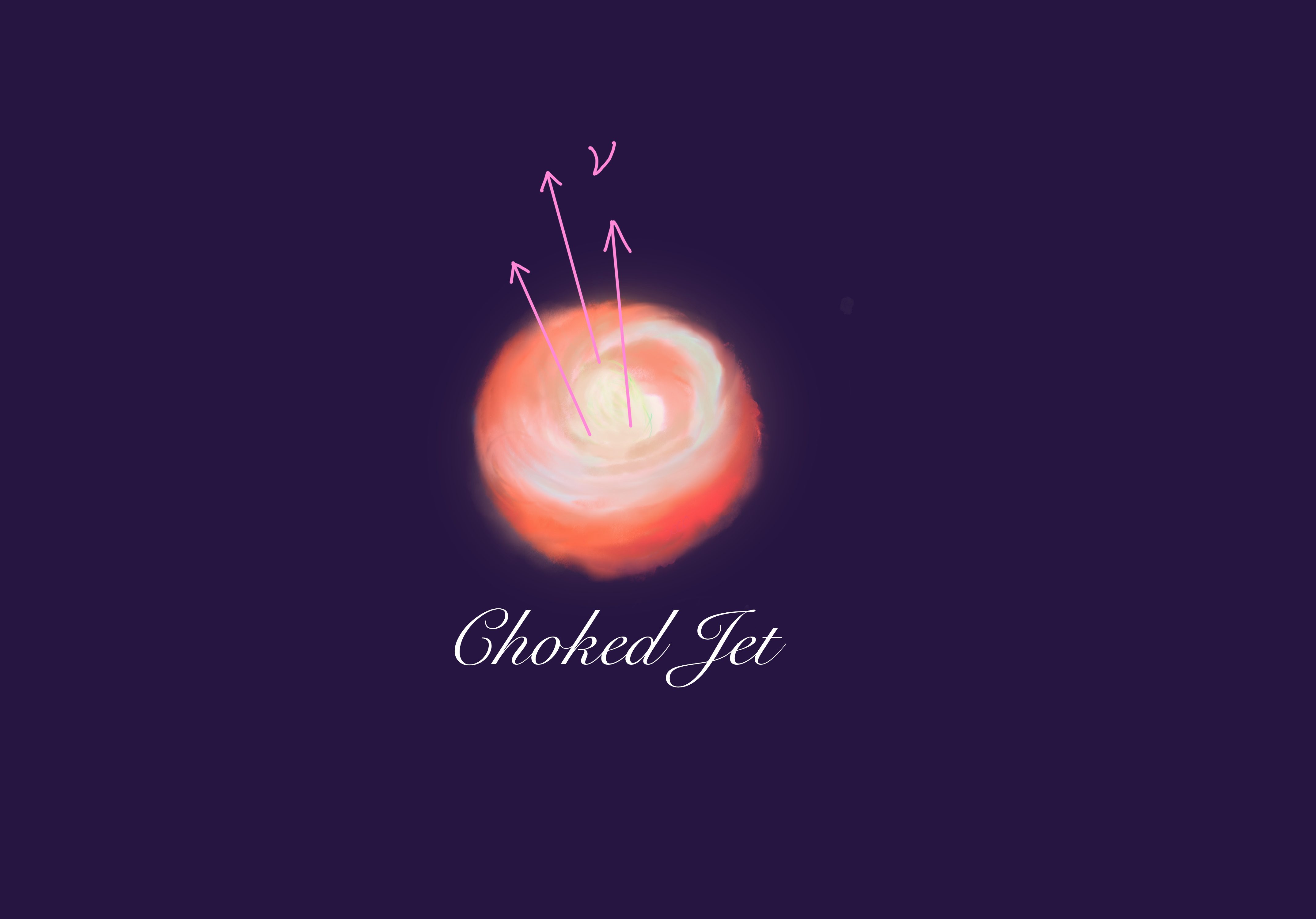}
\includegraphics[width=0.32\linewidth,trim=44cm 23cm 54cm 16cm,clip]{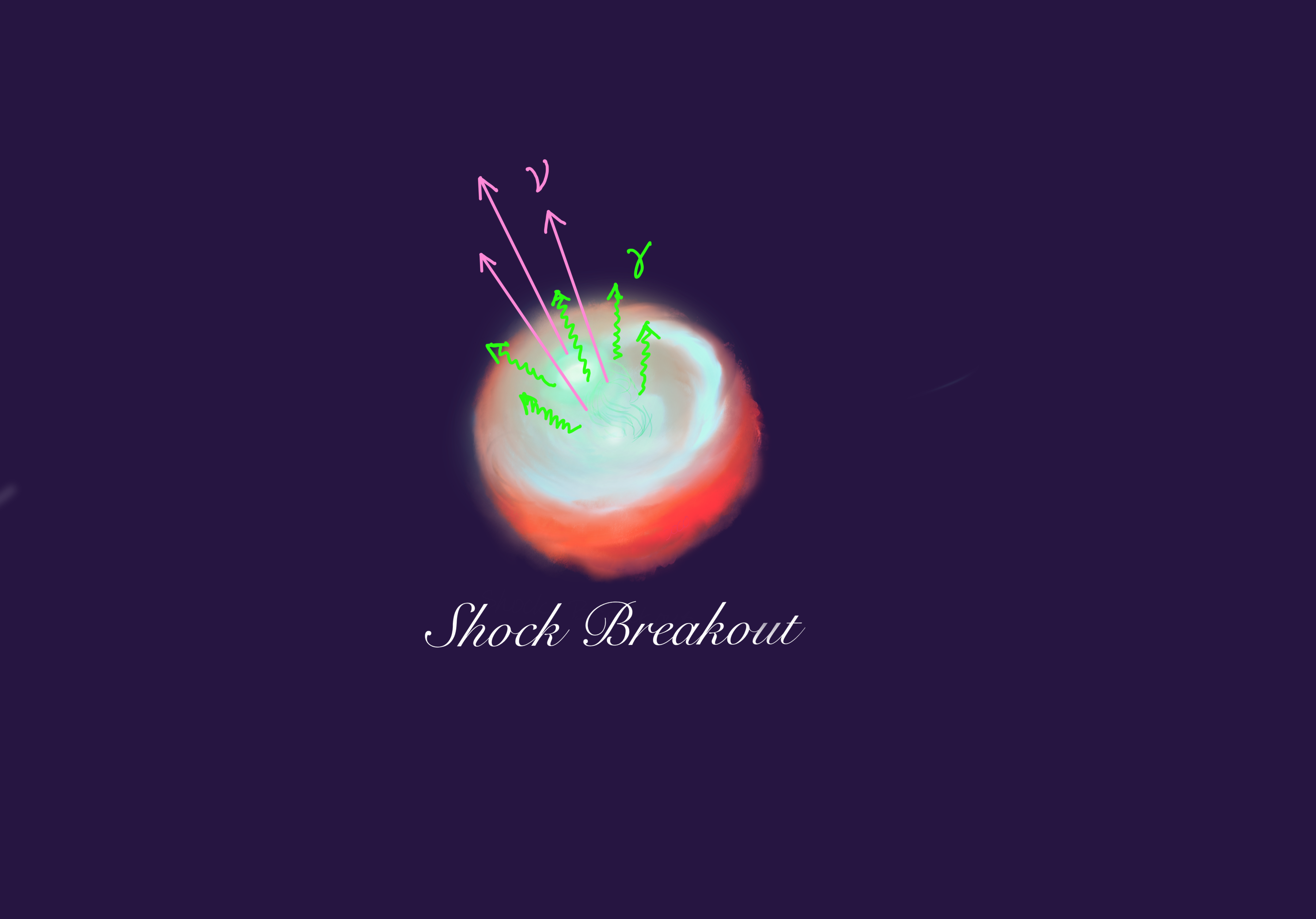}
\includegraphics[width=0.32\linewidth,trim=40cm 19cm 50cm 10cm,clip]{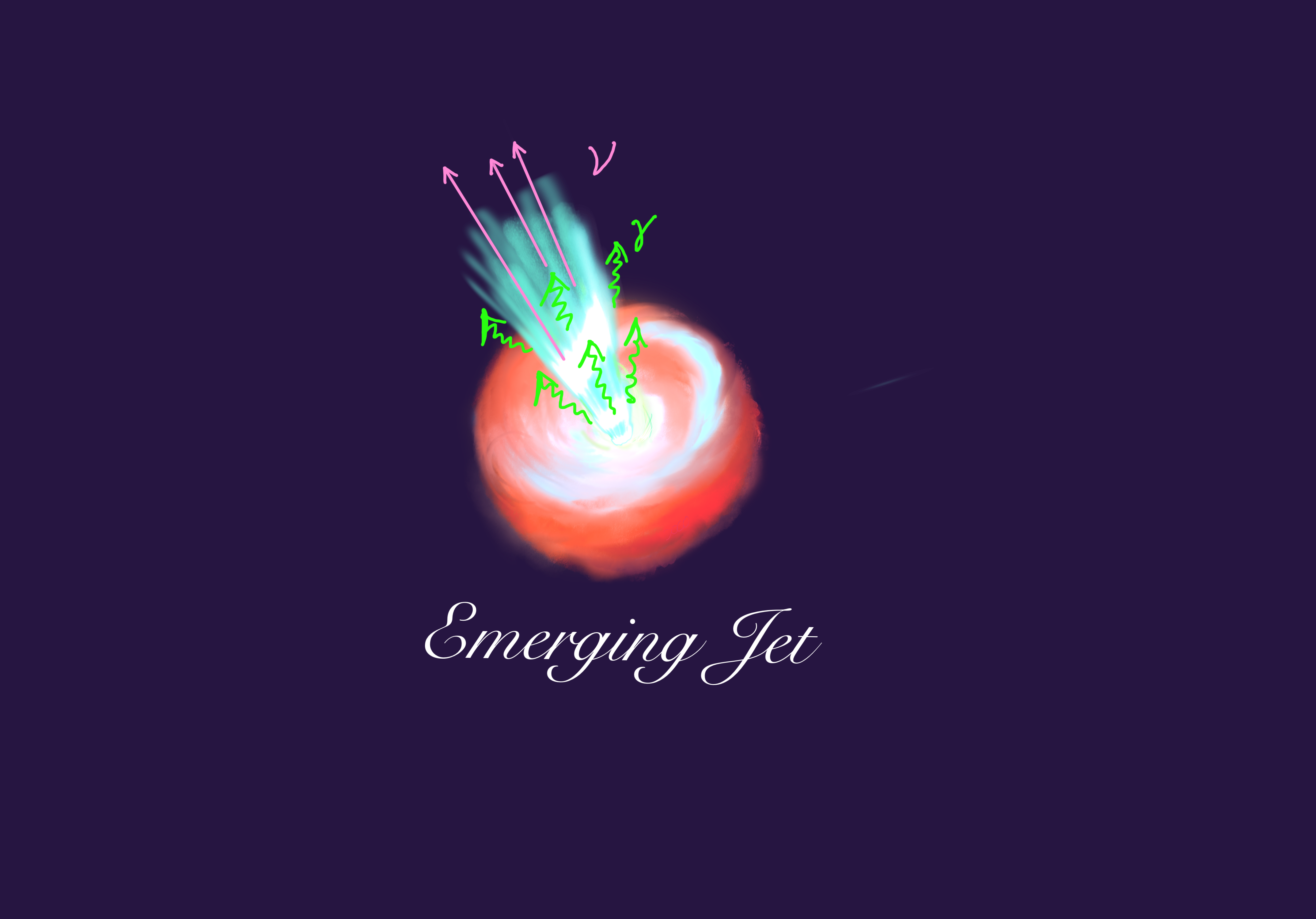}
    \caption{Schematic of massive star deaths. Left: Choked jet model with orphan neutrino emission and no gamma rays. Middle: Shock breakout model for LL~GRBs, producing neutrinos and photons. Right: Emerging jet scenario for typical GRBs and LL~GRBs, with neutrino emission accompanied by photons.}
    \label{fig:LLGRB}
\end{figure*}

Among long GRBs in the Gamma Ray Burst Catalog\footnote{\url{http://heasarc.gsfc.nasa.gov/grbcat/}}, HL GRBs are typically observed at higher redshifts than LL~GRBs. 
This is naturally explained by the intrinsically faint $\gamma$-ray emission of LL~GRBs, which restricts their detectability to the nearby Universe. 
LL~GRBs also appear to constitute a distinct population characterized by higher local event rates~\citep{Liang:2006ci, Sun:2015bda}, mildly relativistic ejecta seen in radio observations~\citep{Soderberg_2004, Soderberg_2006}, and low isotropic-equivalent luminosities during prompt emission~\citep{Campana:2006qe}. Table~\ref{tab:catalog} summarizes the LL~GRBs compiled from the literature~\citep[e.g., ][]{2007ApJ...661L.127M,Sun:2015bda,Zhang:2012jc,DElia:2018xrz,Zhang:2020qbt,Patel_201015A_2023} that satisfy two criteria: isotropic luminosity in the range $10^{46}-10^{50}$ erg/s and preferring models of mild-relativistic jet or breakout shocks. Among the candidates included in our catalog, we focus on the seven most recent events that were observed by \swiftbat{}.
In this section, we briefly review the physical scenarios for the LL~GRBs studied in this work, while our model (Sec.~\ref{sec:model}) is constructed to be sufficiently general to cover the relevant parameter space across them. Our catalog is not intended to represent a complete LL~GRB sample. Sources that failed our selection criteria or had ambiguous categorization (e.g., potential short-GRB origin) or insufficient observational constraints were excluded. 

\begin{table*}[!bth]
\centering
\caption{Selected catalog of low-luminosity GRBs and their observed properties. Columns show the GRB name, associated supernova (SN) if any, Right Ascension (R.A.) and Declination (Dec.) in degrees, trigger time ($T_0$ in MJD), duration of 90\% of the observed $\gamma$-ray photons ($T_{90}$ in seconds, from \swiftbat{} GRBs\footnote{\url{https://swift.gsfc.nasa.gov/results/BATbursts/}} unless cited separately), redshift ($z$), luminosity distance ($d_L$), $\gamma$-ray isotropic-equivalent luminosity $L_{\gamma, \rm iso}$ in 1--10 MeV ($\rm erg\,s^{-1}$), isotropic-equivalent $\gamma$-ray energy $\mathcal{E}_{\gamma, \rm iso}$ ($\rm erg$), and 90\% upper limit on the neutrino flux $\phi_\nu^{\rm 90\%\,UL}$ ($\rm GeV~cm^{-2}$) assuming an $E^{-2}$ spectrum. Unknown values are indicated by --. While multiple scenarios have been proposed to explain LL~GRB prompt emission~\citep{Bromberg:2011fm}, our model (Sec.~\ref{sec:model}) is general and covers the relevant parameter space of all these scenarios.}
\begin{tabular}{l|l|r|r|c|c|c|c|l|l|c}
\hline
LL~GRB & SN & R.A. & Dec. & $T_0$ & $T_{90}$ & $z$ & $d_L$ & log$L_{\gamma, \rm iso}$ & log$\mathcal{E}_{\gamma, \rm iso}$ & $\phi_\nu^{\rm 90\%\,UL}$ \\
& & [deg] & [deg] & [MJD] & [s] & & [Mpc] & [$\rm erg/s$] & [$\rm erg$] & [$\rm GeV~cm^{-2}$] \\
\hline
GRB~980425 & SN 1998bw & 293.7 & -52.8 & 50928.446 & 31~\tnote{a}[a] & 0.0085 & 36 & 46.74 & 47.9 & -- \\
GRB~020903 & SN 2003dh & 342.2 & -19.2 & 52519.92058 & 13~\tnote{b}[b] & 0.25 & - & 48.25~\tnote{c}[c] & 49.40 & -- \\
GRB~031203 & SN 2003lw & 120.6 & 39.8 & 52976.41778 & 30~\tnote{d}[d] & 0.105 & - & 48.58~\tnote{d}[d] & 50.10 & -- \\
GRB~040701 & - & 314.0 & -35.0 & 53190.542303 & 60~\tnote{d}[d] & 0.2146 & - & 48.13~\tnote{d}[d] & 49.90 & -- \\
\textbf{GRB~050826} & - & 80.1 & 20.7 & 53607.76157 & 35.5 & 0.297 & - & 49.1 & 49.9~\tnote{e}[e] & -- \\
\textbf{GRB~060218} & SN 2006aj & 50.4 & 16.9 & 53783.64897 & 2100~\tnote{f} & 0.033 & 145 & 46.0 & 49.3 & -- \\
\textbf{GRB~100316D} & SN 2010bh & 107.6 & -56.3 & 55271.531 & 1300~\tnote{l}[l] & 0.06 & 261 & 46.5 & 49.2 & -- \\
\textbf{GRB~120422A} & SN 2012bz & 136.9 & 14.0 & 56039.300 & 5.4~\tnote{m}[m] & 0.283 & 1464.5 & 48.3 & 50.0 & 0.085~\tnote{h}[h] \\
\textbf{GRB~171205A} & SN 2017iuk & 167.4 & -12.6 & 58092.306 & 190.5 & 0.0368 & 163 & 46.9 & 49.5 & 0.14~\tnote{k}[k] \\
\textbf{GRB~190829A} & SN 2019oyw & 44.5 & -9.0 & 58724.830 & 56.9 & 0.0785 & 358.4 & 48.3 & 50.1 & 0.085~\tnote{k}[k] \\
\textbf{GRB~201015A} & SN 201015A & 354.3 & 53.4 & 59137.952 & 9.8 & 0.426 & 2363.6 & 49.8 & 50.8 & 0.059~\tnote{i}[i] \\
EP 240414a & SN 2024gsa & 191.5 & -9.7 & 60414.4099 & 155~\tnote{j}[j] & 0.401 & 2199.9 & 47.5 & 49.7 & -- \\
\hline
\end{tabular}
\tablecomments{
[a] \cite{Galama:1998ea}, [b] \cite{chandra2012radio}, [c] \cite{Soderberg_2004}, 
[d] \cite{Galama:1998ea}, [e] \cite{2007ApJ...661L.127M}, [f] \cite{2006GCN..4806....1B}, 
[g] \cite{Lucarelli:2022ush}, [h] \cite{2019GCN.25582....1C}, [i] \cite{nu_201015A}, 
[j] \cite{Sun:2024zut}, [k] \cite{IceCube:2020mzw}, [l] \cite{100316D}, [m] \cite{Zhang:2012jc}%
}
\label{tab:catalog}
\end{table*}

Long GRBs are associated with the collapse of massive stars, typically Wolf-Rayet progenitors, during which relativistic jets break out of the stellar surface~\citep{Woosley:1993wj, MacFadyen:1998vz}. 
Associations between long GRBs and supernovae (SNe) provide strong support for the collapsar model~\citep{Galama:1998ea, Campana:2006qe}.
Figure~\ref{fig:LLGRB} illustrates different outcomes after stellar collapse. 
Compared to normal supernova and typical HL GRBs, LL~GRB prompt emission can arise either from the SBO emission~\citep{Campana:2006qe} or internal shocks from a mildly relativistic emerging jet~\citep{Toma:2006iu, Ghisellini:2007ya}. Although the choked-jet scenario predicts minimal radiative output and thus cannot be directly tested through multi-wavelength observations~\citep{Murase:2013ffa, Senno:2015tsn}, studies of LL~GRBs can establish physical limits and offer valuable constraints for future investigations of choked-jet models.

When a jet propagates inside a star, the breakout time ($t_b$) through radius $R_*$ can be estimated from the isotropic-equivalent luminosity of the prompt emission~\citep{Bromberg:2011fm, Bromberg:2011wb},
\begin{align}\label{eq:breakout}
    t_b &\simeq 15\,\epsilon_\gamma \left(\frac{L_{\gamma, \rm iso}}{10^{51}\rm~erg~s^{-1}}\right)^{-1/3} \left(\frac{\theta}{10^\circ}\right)^{2/3} \left(\frac{R_*}{10^{11}\rm~cm} \right)^{2/3} \nonumber \\ &\times \left(\frac{M_*}{15 M_\odot}\right)^{1/3}\rm~s,
\end{align}
where $\epsilon_\gamma$ is the radiation efficiency, $\theta$ is the jet half-opening angle, and $M_*$ is the stellar mass.

A successful jet breakout requires the central engine to remain active longer than the breakout time ($t_{\rm eng}\gtrsim t_b$)~\citep{Bromberg:2011fm}. 
The prompt emission phase, lasting roughly $(t_{\rm eng}-t_b)$, is then powered by dissipation of the jet's kinetic energy, giving rise to the observed non-thermal radiation. In the internal shock model, this emission results from collisions between faster and slower ejecta within the outflow~\citep{Toma:2006iu, Ghisellini:2007ya, Fan:2010br, Irwin:2015rbf}.
If instead the engine activity ceases before breakout ($t_b\gtrsim t_{\rm eng}$), the jet fails to emerge, and the trans-relativistic SBO scenario applies~\citep{Waxman:2007rr, Nakar:2011mq, Bromberg:2011fm, Irwin:2015rbf}.  

Below, we summarize the existing observations and interpretations of LL~GRBs from the catalog (Table~\ref{tab:catalog}), focusing on their connection to either the SBO or IS scenario.

\paragraph{GRB~060218 and GRB~100316D}  
Both exhibit long-duration, smooth prompt emission and soft X-ray components~\citep{Campana:2006qe, Fan:2010br}. 
GRB~060218 is a typical SBO candidate. Its early optical emission and thermal soft X-ray component suggest a mildly relativistic shock propagating through a dense stellar wind~\citep{Campana:2006qe, Waxman:2007rr, Nakar:2015tma}. In this model, the X-rays originate from the shocked wind, while the optical emission arises from the cooler outer layers of the progenitor. However, the SBO interpretation struggles to reproduce the full broadband data: the light-crossing time of $R \sim 5\times 10^{12}\rm~cm$ yields a predicted duration of $\sim 170\rm~s$, much shorter than the observed prompt emission~\citep{Campana:2006qe}. Proposed resolutions include highly aspherical breakout~\citep{Waxman:2007rr} or a jet choked in an extended envelope of radius $\sim 10^{13}-10^{14}\rm~cm$ and mass $\sim 0.01 M_\odot$~\citep{Nakar:2015tma, Irwin:2015rbf}. More recently,~\citep{Irwin:2024aob} considered initial imbalances in gas and radiation pressures and the spectral evolution from broad free-free continuum to quasi-blackbody emission. 

GRB~100316D is among the longest-duration \swiftbat{} GRBs (at least 1300~s) and has a smooth, slowly decaying X-ray light curve similar to that of GRB~060218~\citep{Campana:2006qe,Fan:2010br}. A weak blackbody component contributes a few percent of the total 0.3-10\,keV X-ray flux~\citep{Starling:2010ed}, with large statistical uncertainty~\citep{Margutti:2013pra}.  

The comparable durations, spectral features, and temporal evolution of these two LL~GRBs support interpretations involving trans-relativistic breakout shocks or choked jets within extended stellar envelopes.

\paragraph{GRB~050826, GRB~120422A and GRB~171205A}  
GRB~050826 is categorized as a sub-luminous event that likely contains a wide jet; however, the limited afterglow observations prevent the jet model from being further constrained~\citep{2007ApJ...661L.127M}. GRB~120422A displays a higher $\gamma$-ray luminosity ($L_{\gamma, \rm iso} \sim 10^{49}\rm~erg~s^{-1}$) compared to GRB~060218 and GRB~100316D, with $L_{\gamma, \rm iso} \sim \text{a few} \times 10^{46}\rm~erg~s^{-1}$. Its afterglow observations support the presence of a wide jet with Lorentz factor $\Gamma \sim 6$ with a duration of around 86~s, longer than its $T_{90}$, consistent with a mildly relativistic, successful jet rather than a pure breakout origin~\citep{Zhang:2012jc}.

GRB~171205A is associated with SN 2017iuk, a broad-line type Ic supernova. Early optical spectra reveal extremely fast ejecta ($\sim 115{,}000\rm~km~s^{-1}$), interpreted as a hot cocoon formed by a mildly relativistic jet interacting with the stellar envelope~\citep{Izzo:2019akc}. Its short duration and high peak energy are inconsistent with SBO emission, and the radio afterglow is initially dominated by cocoon emission, with the off-axis jet contributing at later times~\citep{DElia:2018xrz,2024ApJ...962..117L}.  

\paragraph{GRB~190829A and GRB~201015A}
For these two LL~GRBs, there is no clear evidence for an SBO signature.
Instead, early radio observations with the Karl~G.~Jansky Very Large Array (VLA) and the Australia Telescope Compact Array (ATCA) of GRB~190829A suggest either a mildly relativistic outflow or a structured jet observed slightly off-axis. GRB~190829A is notable for its very-high-energy (VHE) $\gamma$-ray afterglow detected by the High Energy Stereoscopic System (H.E.S.S.)~\citep{HESS:2021dbz}, while multi-wavelength follow-up observations constrain the jet Lorentz factor to $\Gamma \lesssim 40$~\citep{2023ApJ...952..127Z}. A supernova was clearly detected in the optical follow-up~\citep{Suda:2021dox, Giarratana:2022wtt}. 

GRB~201015A similarly shows no evidence of breakout shocks. Its prompt emission and afterglow are consistent with a wider, intrinsically low-luminosity jet with a half-opening angle $\theta_j \gtrsim 16^\circ$~\citep{Patel:2023lkd}, with the jet Lorentz factor inferred to be moderate. 

\paragraph{EP~240414a}  
The discovery of the fast X-ray transient EP~240414a by the Einstein Probe (EP) mission has advanced the search for high-energy transients associated with the deaths of massive stars~\citep{Sun:2024zut}.  
EP~240414a is associated with the Type Ic-BL supernova SN 2024gsa at redshift $z = 0.401$. Its X-ray spectrum is very soft ($E_{\rm peak} < 1.3\rm~keV$), consistent with emission produced by the interaction of a weak relativistic jet within a dense shell surrounding the progenitor~\citep{Sun:2024zut,Bright:2024zaw, Hamidani:2025nup,Zheng:2025yhv}.  
Modeling of the early optical and X-ray data suggests a successful but low-power jet with an initial Lorentz factor $\Gamma_0 \gtrsim 13$~\citep{Sun:2024zut}.

Together, these observations illustrate that LL~GRBs likely span a continuum between failed or choked jets producing breakout shocks and successful, mildly relativistic jets that break out of the stellar envelope. To further understand their physical properties, joint modeling of multiwavelength and neutrino data is essential.  
In this work, we perform a comprehensive analysis of the seven selected LL~GRBs with prompt emission observed by \swiftbat{} and/or \fermilat{}, incorporating electromagnetic observations and upper limits to constrain their physical origins.

\section{High-energy neutrinos and gamma rays from LL~GRBs}\label{sec:model}
We construct a unified model that incorporates the essential physical processes -- particle acceleration, interactions, and energy redistribution via cooling -- while allowing model parameters to span ranges appropriate for both IS and SBO. We introduce a set of free and fixed parameters to describe the physical conditions in the shock region, guided by empirical and theoretical inferences. 
We will first discuss the differences between IS and SBO in our model consideration. Then, their common physical processes that are employed in our unified model are discussed in detail. 

Internal shocks arise when fast-moving ejecta collide with slower precursors, while SBO occurs when a mildly relativistic shock emerges from the progenitor's envelope~\citep{1994ApJ...430L..93R,Kobayashi1997,DaigneMochkovitch1998,Campana:2006qe,Soderberg_2006}. These scenarios are primarily distinguished by the bulk Lorentz factor of the outflow ($\Gamma$) and the radius of the energy dissipation region ($R$). Accordingly, our unified model keeps these parameters free over a broad range to accommodate both cases.

The dissipation radius $R$ determines the density of ambient photons, which serve as targets for $p\gamma$ interactions. The efficiencies of energy dissipation -- into magnetic fields, heating of the ambient medium, and non-thermal particle acceleration -- depend on $R$~\citep{Peer:2015eek}. For internal shocks, $R \sim 2 \Gamma^2 \delta t_{\rm min} \sim 6 \times 10^{15} (\Gamma / 10)(\delta t_{\rm min}/10^3~\rm s)$\,cm~\citep{Kobayashi1997,DaigneMochkovitch1998}, while SBO shocks, with $\Gamma \sim 1$-3 and $\delta t_{\rm min} \sim 100$-1000\,s, typically occur at smaller radii $R \sim 10^{12}$-$10^{13}$\,cm~\citep{Campana:2006qe,Soderberg_2006}. 
We allow $R$ to range from $10^{11.5}$\,cm to $10^{16}$\,cm, and $\Gamma_j$ in the range of [2, 20], covering both internal shock and SBO scenarios~\citep{Kumar:2014upa}.

Because LL~GRB prompt emission data are limited in both statistics and energy coverage (Sec.~\ref{sec:mm}), we do not model individual shocks or pulse variability. Instead, we consider a time-averaged burst, represented by a homogeneous shell expanding at $\beta = \sqrt{1-1/\Gamma^2}$ with co-moving width $\Delta \simeq R/(\Gamma \beta)$.

Cosmic-ray protons and electrons are assumed to undergo diffusive shock acceleration (DSA)~\citep{Drury:1983zz, Blandford:1987pw, Sironi:2015oza}. Electrons dominate the target photon field via synchrotron and inverse-Compton (IC) emission, including secondary electrons from hadronic processes. The injected particle spectra follow a power-law with exponential cutoff:
\begin{equation}
\frac{dn}{d\varepsilon_i^\prime} \propto {\varepsilon_i^\prime}^{-s_i} \exp\left(-\frac{\varepsilon_i^\prime}{\varepsilon^\prime_{i, \rm cutoff}}\right),
\end{equation}
where $i = p, e$. We adopt $s_p = 2$ for CR protons and let $s_e$ vary between 2 and 3 to match the observed photon spectra. The particle energy densities are related to the radiation energy density by $U_i = \xi_i U_\gamma$, with $U_\gamma \simeq L_{\gamma, \rm iso}/(4\pi R^2 \Gamma^2 c)$. The minimum proton energy is taken as $\varepsilon_{p,\rm min}^\prime \sim 10 m_p c^2$ to be compatible with mildly relativistic shocks.

The total energy density of electrons is
\begin{equation}
u_e \;\approx\; n_e \,\gamma_{e,\min} m_e c^2 \,\frac{s_e-1}{s_e-2},
\end{equation}
valid for $s_e > 2$, where $n_e$ is the electron number density and $\gamma_{e,\rm min}$ is the minimum Lorentz factor of injected electrons. Since electrons and protons share the dissipated kinetic energy of the shocks, we introduce a partition parameter $\epsilon_e$ representing the fraction of shock energy allocated to accelerated electrons:
\begin{equation}
u_e \;\sim\; \epsilon_e\, n_p (\Gamma_{\rm shock}-1) m_p c^2,
\end{equation}
which yields
\begin{equation}
\gamma_{e,\min} \;\sim\; \epsilon_e\, \frac{s_e-2}{s_e-1}\, \frac{m_p}{m_e}\, (\Gamma_{\rm shock}-1),
\end{equation}
assuming charge neutrality ($n_e = n_p$). For $s_e = 2$, $\gamma_{e,\rm min}$ satisfies a transcendental equation, giving an approximate analytical solution. 
For LL~GRBs, the electron distribution is typically in the fast-cooling regime, and observational coverage below $\sim 15$\,keV is limited. Therefore, current data cannot meaningfully constrain $\gamma_{e,\rm min}$. To account for this and the uncertainty in the effective shock Lorentz factor, $\Gamma_{\rm shock}$, we fit for $\epsilon_e$ instead, allowing it to vary over a broad range [1\%, 30\%]\footnote{Since $\epsilon_e$ and $\Gamma_{\rm shell}$ together determine $s_e$, to estimate the value of $\gamma_{e, \rm min}$, the choice of nominal values and allowed range primarily affects theoretical interpretation of the shock dynamics, not the fitted flux, which is the focus of this study.}, consistent with previous studies~\citep{Zhang_2002}.
The value of $s_e$ is free between 2 and 3 and can be constrained by X-ray and $\gamma$-ray spectra during the prompt emission (Sec.~\ref{sec:fit_model}). Consequently, $\gamma_{e,\rm min}$ contributes to the observed multi-wavelength flux but is degenerate with $s_e$.

The maximum energy of a charged particle is set by acceleration and cooling: $\varepsilon_{\rm cutoff} \approx \eta_{\rm acc}^{-1} eB^\prime R/\Gamma$, with acceleration efficiency $\eta_{\rm acc}=10$ in our study~\citep{Sironi:2015oza}, and the comoving magnetic field estimated as:
\begin{equation}
B' \simeq 220~{\rm G}~\xi_B^{1/2} \left(\frac{L_{\gamma, \rm iso}}{10^{47}~\rm erg~s^{-1}}\right)^{1/2} \left(\frac{R}{10^{15}~\rm cm}\right)^{-1} \left(\frac{\Gamma}{10}\right)^{-1},
\end{equation}\label{eq:B}
with $\xi_B$ representing the fraction of radiation energy in magnetic fields. 
The energy partitions between electrons, protons, and magnetic fields are characterized by the parameters $\xi_e$, $\xi_p$, and $\xi_B$. We allow $\xi_p$ to vary between $10^{-2}$ and $10^{2}$, $\xi_e$ in $10^{-0.5}$ and $10^{2}$, and $\xi_B$ between $10^{-3}$ and $10^{3}$, reflecting the wide theoretical uncertainty in shock microphysics.

These parameter ranges define the priors used in the subsequent MCMC fitting (Sec.~\ref{sec:fit_model}) and are summarized in Table~\ref{tab:best_fit_params} along with the fitting results. They are chosen to cover the physically plausible regime for LL~GRBs while ensuring the model remains flexible enough to accommodate the observed multi-wavelength data.

The relevant timescales of acceleration, dynamical expansion, and cooling are balanced as:
\begin{align}
    t_{\rm acc} & \approx \eta_{\rm acc} r_L/c \approx \eta_{\rm acc} \varepsilon_p^\prime/ZeB^\prime,\\
    t_{\rm dyn} & \approx t_{\rm ad}\approx R/\Gamma \beta c,\\
    t_{\rm loss}^{-1} &= t_{\rm syn}^{-1} + t_{p\gamma-\rm meson}^{-1} + t_{\rm BH-pair}^{-1},
\end{align}\label{eq:timescales}
where $t_{\rm loss}^{-1}$ is the total energy loss rate for charged particles, and $t_{\rm syn}$, $t_{p\gamma-{\rm meson}}$, and $t_{\rm BH-pair}$ denote synchrotron, photomeson ($p\gamma$-meson), and Bethe-Heitler (BH) pair-production cooling timescales, respectively. The adiabatic cooling timescale, $t_{\rm ad}$, approximates the dynamical expansion timescale, $t_{\rm dyn}$. 
Photomeson interactions ($p + \gamma \rightarrow \pi^{\pm,0} + {\rm hadrons}$) produce secondary pions that decay into muons, neutrinos, and $\gamma$ rays. In our model, $p\gamma$ interactions are the dominant channel for meson production during the prompt emission phase, while $pp$ interactions -- though capable of generating similar secondaries -- are subdominant and thus neglected. The maximum energies of the secondary pions and muons are determined by the balance between their decay and synchrotron cooling timescales, which sets the high-energy cutoff of the neutrino spectrum (see the corresponding timescales in Figs.~\ref{fig:17_05}--\ref{fig:06_10_12} and discussion in Sec.~\ref{sec:result_discussion}).

The all-flavor neutrino spectrum in the observer frame is:
\begin{equation}
\varepsilon_\nu^\prime Q_{\varepsilon_\nu}^\prime \approx \frac{3}{8} f_{\rm sup} f_{p\gamma} \varepsilon_p^\prime Q_{\varepsilon_p}^\prime,
\end{equation}
where $\varepsilon_\nu^\prime \sim 0.05 \varepsilon_{\rm p}^\prime$ is the neutrino energy in the comoving frame~\citep{Waxman:1997ti}, and $\varepsilon_p^\prime Q_{\varepsilon_p}^\prime$ is the injection rate of CR protons.
The suppression factor, $f_{\rm sup} \simeq f_{\rm sup, \pi} f_{\rm sup, \mu}$~\citep{Kimura:2022zyg}, accounts for the energy cooling of the primary pions and secondary muons, primarily via synchrotron radiation. Each component can be estimated as $f_{\rm sup, i} = 1 - {\rm exp} (-t_{i, \rm cool} / t_{i, \rm decay})$, where $i$={$\pi$, $\mu$}, $t_{i, \rm decay}$ is the decay timescale, and $t_{i, \rm cool}$ is the cooling timescale.
We solve the transport equation of the comoving-frame proton number density, $n_{\varepsilon_p^\prime}^{p}$:
\begin{align}
\frac{\partial n_{\varepsilon_p^\prime}^{p}}{\partial t} 
&= \overbrace{\dot{n}_{\varepsilon_p^\prime}^{\rm inj}}^{\text{injection}} 
 + \overbrace{\frac{\partial}{\partial t} (n_{\varepsilon_p^\prime}^{p\gamma} + n_{\varepsilon_p^\prime}^{\beta_{\rm dec}})}^{\text{secondary produced}} \\
&\quad - \underbrace{\frac{n_{\varepsilon_p^\prime}^{p}}{t_{\rm esc}} 
          - \frac{n_{\varepsilon_p^\prime}^{p}}{t_{p\gamma}}}_{\text{escape and $p\gamma$-meson loss}} \\
&\quad - \underbrace{\frac{\partial}{\partial \varepsilon_p^\prime} 
     \Big[(P_{\rm ad}^p + P_{\rm BH-pair}^p + P_{\rm syn}^p) n_{\varepsilon_p^\prime}^{p}\Big]}_{\text{energy redistribution}},
\end{align}
with terms representing injection, secondary production, escape, $p\gamma$-meson losses, and energy redistribution via adiabatic, synchrotron, and Bethe-Heitler pair production cooling. 
Solving this yields the neutrino production rate from pion and muon decays~\citep{Lipari:2007su}, and the $\gamma$-ray spectrum is related by:
\begin{equation}
  \varepsilon_\gamma^\prime Q_{\varepsilon_\gamma^\prime}  \sim \frac{4}{3K} \varepsilon_\nu^\prime Q_{\varepsilon_\nu^\prime}  |_{\varepsilon_\gamma^\prime = 2 \varepsilon_\nu^\prime},
\end{equation}
with $K=1$ for $p\gamma$ interactions~\citep{Murase:2015xka}.
Photons and electrons produced by secondary decays, ambient photon interactions, and leptonic processes undergo electromagnetic cascades, redistributing energy and driving the system toward a steady state, where particle injection, energy losses, and escape balance and the overall distributions remain approximately constant over time. The transport equations for photons and electrons are given by
\begin{align}\label{eq:transport-photon}
\frac{\partial n_{\varepsilon_\gamma^\prime}^{\gamma}}{\partial t} &= \overbrace{\dot{n}_{\varepsilon_\gamma^\prime}^{\rm inj}}^{\text{injection}}+ \overbrace{\frac{\partial}{\partial t} \left(n_{\varepsilon_\gamma^\prime}^{\rm syn} + n_{\varepsilon_\gamma^\prime}^{\rm IC} + n_{\varepsilon_\gamma^\prime}^{\rm p\gamma}\right)}^{\text{produced}}\\
&-\underbrace{n_{\varepsilon_\gamma^\prime}^{\gamma} \left(\frac{1}{t_{\rm esc}^\gamma} + \frac{1}{t_{\gamma \gamma}}\right) \nonumber}_{\text{escape and annihilation loss}},
\end{align}
and
\begin{align}\label{eq:transport-electron}
    \frac{\partial n_{\varepsilon_e^\prime}^{e}}{\partial t} &= \overbrace{\dot{n}_{\varepsilon_e^\prime}^{\rm inj}}^{\text{injection}} + \overbrace{\frac{\partial}{\partial t} \left(n_{\varepsilon_e^\prime}^{\gamma \gamma} + n_{\varepsilon_e^\prime}^{\rm BH-pair} + n_{\varepsilon_e^\prime}^{\rm p\gamma} + n_{\varepsilon_e^\prime}^{\beta_{\rm dec}}\right)}^{\text{produced}}\\
    &-\underbrace{\frac{n_{\varepsilon_e^\prime}^{e}}{t_{\rm esc}^e}}_{\text{escape loss}}
    - \underbrace{\frac{\partial}{\partial \varepsilon_e^\prime} \left[(P_{\rm ad}^e + P_{\rm syn}^e + P_{\rm IC}^e) n_{\varepsilon_e^\prime}^{e} \right] \nonumber}_{\rm energy\ redistribution}.
\end{align}
Photon escape timescale is $t_{\rm esc}^\gamma \equiv \Delta / c$, $t_{\gamma \gamma}$ is the $\gamma\gamma$ annihilation timescale, and $\dot{n}_{\varepsilon\gamma}^{\rm inj} = 0$ (no primary photons). Electrons are injected at $\dot{n}_{\varepsilon_e}^{\rm inj} = (1/t_{\rm dyn}) dn/d\varepsilon_e^\prime$ and also produced via $\gamma\gamma$, Bethe-Heitler pair production, $p\gamma$-meson, and $\beta$-decay processes.

While solving these transport equations, the target photon field is updated on each dynamical timestep $t_{\rm dyn}$, and the system reaches steady state when consecutive changes in photon density become negligible~\citep{Zhang:2023ewt}. In our model, the escape timescale of charged particles is negligible compared to the cooling timescale.

\section{Multi-messenger Data and Fit} \label{sec:mm}
We analyze seven more recent LL~GRBs listed in Table~\ref{tab:catalog} and observed by \swiftbat{}. Whenever available, X-ray and $\gamma$-ray data during the prompt emission phase are used for each GRB. The proton energy density and the key physical processes of electromagnetic radiation and leptohadronic interactions are modeled with the numerical code, Astrophysical Multimessenger Emission Simulator ({\sc AMES})~\citep{Murase_2018PhRvD..97h1301M, Zhang:2023ewt}. The choice of free and fixed model parameters has been discussed in Sec.~\ref{sec:model}. Here, we detail how the observational data and multi-wavelength measurements are incorporated to constrain the physical parameters of our model for each LL~GRBs.

The observed properties of individual LL~GRBs, such as redshift, isotropic-equivalent $\gamma$-ray luminosity ($L_{\gamma, \rm iso}$), and duration of $T_{90}$, etc, are distinct and used as input parameters for the model fitting. The radius ($R$) of the energy dissipation region is treated as a free parameter, as are the parameters describing CR proton and electron acceleration.
The $\gamma$-ray emission computed with {\sc AMES} is fitted to the observed X-ray and $\gamma$-ray SEDs during the prompt phase, as measured by \swiftbat{} and \fermilat{}.
For four LL~GRBs in our sample, 90\% confidence-level (C.L.) upper limits on the observed neutrino fluxes are found ~\citep{IceCube:2020mzw,nu_201015A,2019GCN.25582....1C}, but they are too high to impose meaningful constraints on the model parameters. These limits are therefore not used in the fits, although they are listed in Table~\ref{tab:catalog} for completeness. Further details of the multiwavelength data analysis are provided next.

\subsection{\swiftbat{} and \fermilat{} Data Analyses}
We use the observed energy spectra during T$_{90}$ to study prompt emission. 
Official Pulse Height Analysis (PHA) products for each GRB are obtained from the \swiftbat{} GRB catalog (batgrbcat)\footnote{\url{https://swift.gsfc.nasa.gov/results/batgrbcat/}}~\citep{Lien_2016} and analyzed with {\sc XSPEC}~\citep{1996ASPC..101...17A} (version 12.15.0). The details of the archival data from \swiftbat{} are summarized in Table~\ref{tab:obsdata} in Appendix~\ref{appendix:x-ray}. Response matrices are generated using {\sc batdrmgen}. In addition to using the archival SEDs in our model fits, we perform independent spectral fits using a power-law model with an exponential cutoff for cross-validation. The resulting best-fit parameters, summarized in Table~\ref{tab:x-ray_analysis}, are consistent with those reported in previous analyses of these GRBs.

To obtain $\gamma$-ray upper limits during the prompt emission, we analyze Pass8 \textit{Fermi}-LAT data using \textit{fermipy} (v1.3.1)\footnote{\url{https://fermi.gsfc.nasa.gov/cgi-bin/ssc/LAT/LATDataQuery.cgi}}. Among our sample, only four LL~GRBs -- GRB~100316D, GRB~120422A, GRB~171205A, and GRB~190829A -- were within the \fermilat{} field of view during their prompt emission or shortly after the trigger, $T_0$. For GRB~100316D and GRB~190829A, we extract LAT data within the duration T$_{90}$ (see Table~\ref{tab:catalog}). For GRB~171205A and GRB~201015A, where the \fermilat{} observations started slightly later after trigger, we use the interval [T$_0$, +1000\,s] to derive time-averaged upper limits close to the prompt phase.
For GRB~060218, which occurred before the launch of \fermilat{}, and GRB~120422A, which lacked \fermilat{} observations within the first 5,000\,s after T$_0$, no $\gamma$-ray upper limits are included in the fitting. The corresponding \fermilat{} time ranges are also included in Table~\ref{tab:obsdata}, alongside the \swiftbat{} archival data. We selected photons in the 50\,MeV-300\,GeV band within a $20^\circ$ region of interest (ROI) centered on the GRB positions. Events were binned logarithmically at [$\log_{10}(50)$, 2, 2.5, 3, 4, 5] for spectral fitting.
The following \fermilat{} transient-source selections are applied: $evclass=128$, $evtype=3$, $\zeta_{max}=105^\circ$, and {\sc DATA\_QUAL}$>0$ \& {\sc LAT\_CONFIG}==1.
The Galactic diffuse and isotropic backgrounds were modeled by {\sc gll\_iem\_v07} and {\sc iso\_P8R3\_TRANSIENT020\_V3\_v1}, respectively.

We first freed the parameters of all \textit{Fermi}-4FGL~\cite{Abdollahi_2020} sources within $10^\circ$ of the target LL~GRB, along with the normalizations of the Galactic and isotropic backgrounds and the spectral parameters of the LL~GRB, assuming a power-law model. 
A likelihood fit was then performed, and all the target LL~GRBs returned $TS=0$, indicating no significant detection. Consequently, 95\% confidence upper limits were derived using the standard Fermi-LAT SED analysis, assuming a power-law spectrum with a fixed photon index of 2 while keeping the background parameters free in each energy-bin-independent fit. The resulting $\gamma$-ray 95\% upper limits for the four LL~GRBs are shown in Fig.~\ref{fig:17_05}-\ref{fig:06_10_12}.

\subsection{Fitting with leptohadronic model}\label{sec:fit_model}
We use the Markov Chain Monte Carlo (MCMC) method implemented in {\sc emcee} (v3.1.6) to estimate the uncertainty ranges of the model parameters. The \swiftbat{} observed spectra and the 95\% C.L.\ upper limits of the \fermilat{} $\gamma$-ray fluxes are fitted simultaneously. Their respective log-likelihoods are defined in Eqs.~\ref{eq:llh_xray} and~\ref{eq:prob_gamma}. For the observed X-ray spectra, we adopt the conventional log-likelihood equation:
\begin{equation}
-2{\rm ln} \mathcal{L}_{{\gamma}} = \sum_{i} \left[{\frac{(n_{\text{exp},i} - n_{\text{obs},i})^2}{\sigma^2_{\text{exp},i}} +\ln(\sigma^2_{\text{exp},i})}\right]\,
\label{eq:llh_xray}
\end{equation}
where $i$ denotes the $i$-th spectral bin, $n_{\text{exp},i}$ and $n_{\text{obs},i}$ are the predicted and observed photon counts, and $\sigma_{\text{exp},i}$ is the corresponding statistical uncertainty. 
For bins where the best-fit flux is consistent with zero signal, the \fermilat{}SED analysis reports a $95\%$ C.L.\ upper limit.
In such cases, we impose no penalty as long as the model prediction remains below the reported upper limit (UL); otherwise, the fit is penalized as:
\begin{equation}
-2 \ln \mathcal{L}_{\gamma, \rm UL} \;=\sum_{i} \;
\begin{cases}
0, & n_{\rm exp,i} \le {\rm UL}_i , \\[6pt]
\displaystyle \frac{\big(n_{\rm exp,i})^2}{\left(\sigma_i^{\rm UL}\right)^2} ,
& n_{\rm exp,i} > {\rm UL}_i
\end{cases}
\label{eq:prob_gamma}
\end{equation}
where $\sigma_i^{\rm UL} = {\rm UL}_i / 1.64$ is the $1\sigma$ equivalent standard deviation, assuming a Gaussian distribution centered at zero signal.
Since the reported neutrino upper limits exceed the model-predicted fluxes for the LL~GRBs, they are not used to constrain the fits. They are included in Table~\ref{tab:catalog} for completeness and reference. 

Following the methodology described above, we fit our model to the X-ray and $\gamma$-ray SEDs of each LL~GRB individually. The posterior median parameter values and their 1$\sigma$ uncertainties are derived from the converged MCMC chains after discarding the burn-in phase. The results are discussed in the next section.

\section{Results and Discussion}\label{sec:result_discussion}

\begin{table*}[!tbh]
\centering
\caption{Posterior medians with 1\,$\sigma$ uncertainties (shown in brackets, i.e., [lower bound, upper bound]) of physical parameters for the prompt emission modeling of selected LL~GRBs, constrained by multi-wavelength data.}
\label{tab:best_fit_params}
\begin{tabular}{lcccc}
\hline
Parameter & 050826 & 060218 & 100316D & 120422A \\
\hline
$\log(\xi_B)$ & $-2.3\,[-3.02, -1.82]$ & $1.6\,[0.92,2.99]$ & $2.0\,[1.52,3.01]$ & $1.0\,[0.06,2.96]$ \\
$\log(\xi_e)$ & $-0.2\,[-0.39,-0.05]$ & $1.0\,[0.72,1.30]$ & $0.7\,[0.33,1.04]$ & $0.5\,[0.22,0.84]$ \\
$\log(\xi_p)$ & $-0.7\,[-1.63,0.10]$ & $0.2\,[-0.59,2.00]$ & $0.1\,[-0.78,1.77]$ & $0.1\,[-0.84,1.59]$ \\
$\log(R~[\mathrm{cm}])$ & $13.4\,[11.52,14.41]$ & $13.6\,[11.51,14.54]$ & $12.8\,[11.51,13.54]$ & $13.6\,[11.52,14.44]$ \\
$\Gamma_j$ & $13.3\,[10.29,19.96]$ & $10.6\,[2.07,14.01]$ & $8.8\,[2.03,12.10]$ & $11.0\,[2.04,14.29]$ \\
$s_e$ & $2.6\,[2.41,3.03]$ & $2.5\,[2.12,2.77]$ & $2.5\,[2.24,2.82]$ & $2.5\,[2.04,2.70]$ \\
$\epsilon_e$ & $0.2\,[0.14,0.34]$ & $0.2\,[0.15,0.34]$ & $0.2\,[0.07,0.26]$ & $0.2\,[0.15,0.34]$ \\
\hline
\end{tabular}
\end{table*}

\begin{table*}[!tbh]
\centering
\addtocounter{table}{-1} 
\caption{(Continued) Posterior medians with 1\,$\sigma$ uncertainties of physical parameters for the remaining LL~GRBs. The last column shows the allowed prior ranges used in the MCMC fitting.}
\begin{tabular}{lcccc}
\hline
Parameter & 171205A & 190829A & 201015A & Fitting range \\
\hline
$\log(\xi_B)$ & $-2.7\,[-3.00, -2.53]$ & $2.1\,[1.70,3.00]$ & $1.8\,[1.27,3.00]$ & $[-3,3]$ \\
$\log(\xi_e)$ & $0.3\,[0.16,0.48]$ & $0.6\,[0.36,0.77]$ & $0.2\,[0.00,0.46]$ & $[-0.5,2]$ \\
$\log(\xi_p)$ & $-0.2\,[-0.93,1.13]$ & $-0.6\,[-1.99,-0.06]$ & $-0.6\,[-2.03,-0.06]$ & $[-2,2]$ \\
$\log(R~[\mathrm{cm}])$ & $11.9\,[11.53,12.12]$ & $13.0\,[12.06,13.73]$ & $14.0\,[12.75,15.25]$ & $[11.5,16]$ \\
$\Gamma_j$ & $12.8\,[9.31,18.96]$ & $8.9\,[2.02,12.02]$ & $11.0\,[5.49,17.57]$ & $[2,20]$ \\
$s_e$ & $2.5\,[2.32,2.98]$ & $2.8\,[2.71,3.00]$ & $2.7\,[2.54,3.01]$ & $[2,3]$ \\
$\epsilon_e$ & $0.1\,[0.03,0.16]$ & $0.2\,[0.09,0.27]$ & $0.2\,[0.07,0.26]$ & $[0.01,0.3]$ \\
\hline
\end{tabular}
\tablecomments{Columns correspond to GRB names, while rows list the physical parameters: logarithm of the magnetic energy ratio ($\log\xi_B$), logarithm of the electron energy ratio ($\log\xi_e$), logarithm of the CR loading factor ($\log\xi_p$), logarithm of the emission radius in unit of cm ($\log(R~[\mathrm{cm}])$), jet Lorentz factor ($\Gamma_j$), electron spectral index ($s_e$), and fraction of shock energy transferred to accelerated electrons ($\epsilon_e$).}
\end{table*}

\begin{figure*}[!tbh]
\centering
\includegraphics[width=0.32\linewidth]{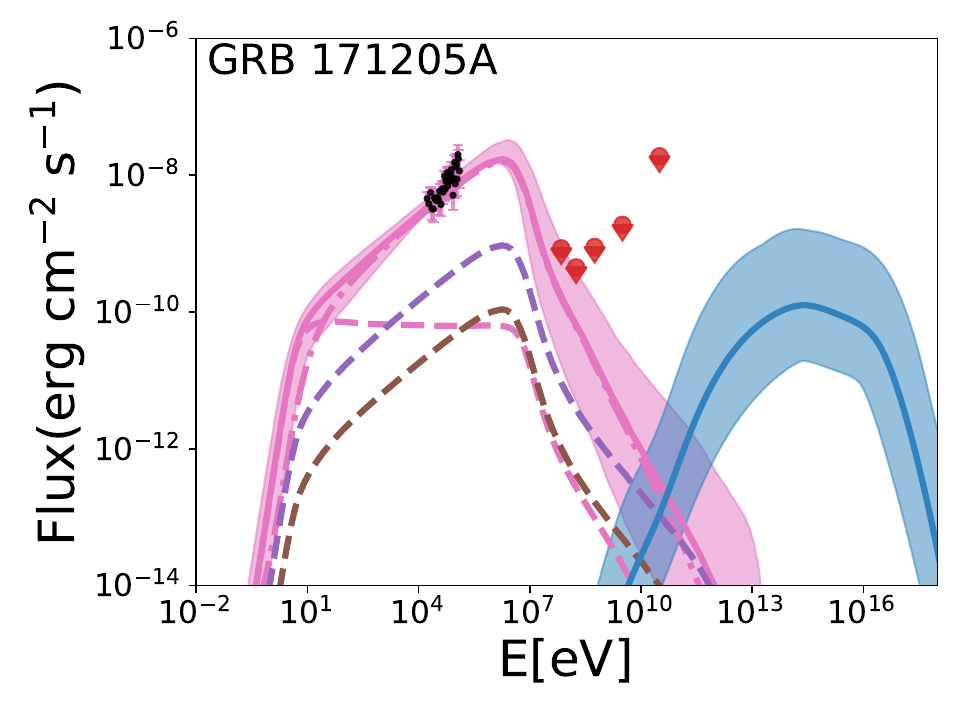}
\includegraphics[width=0.32\linewidth]{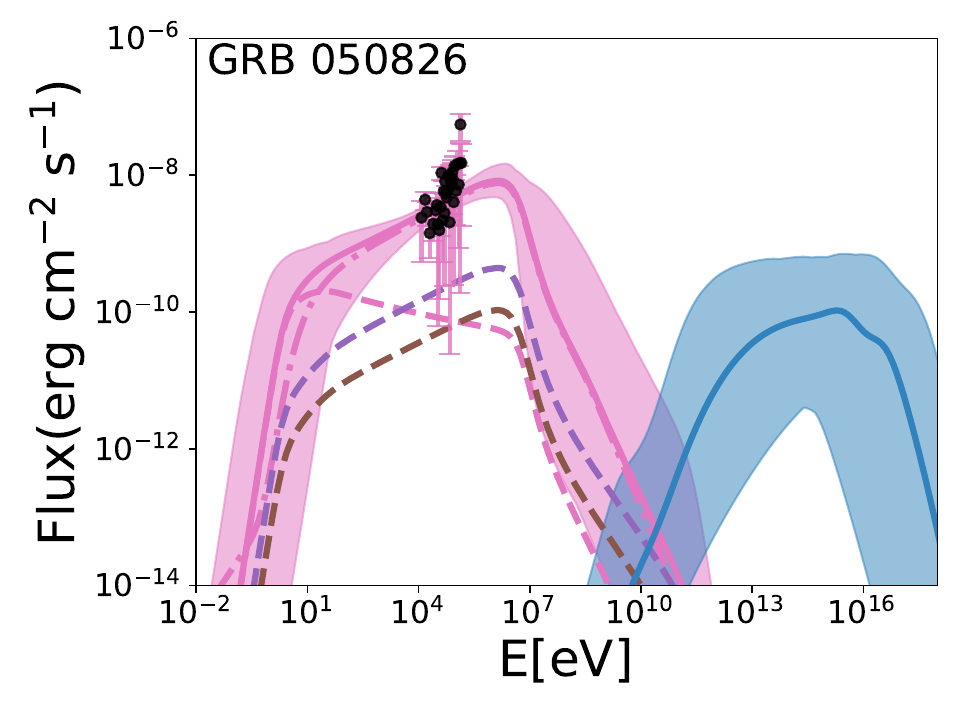}
\includegraphics[width=0.32\linewidth,trim=2cm 1cm 2cm 2cm,clip]{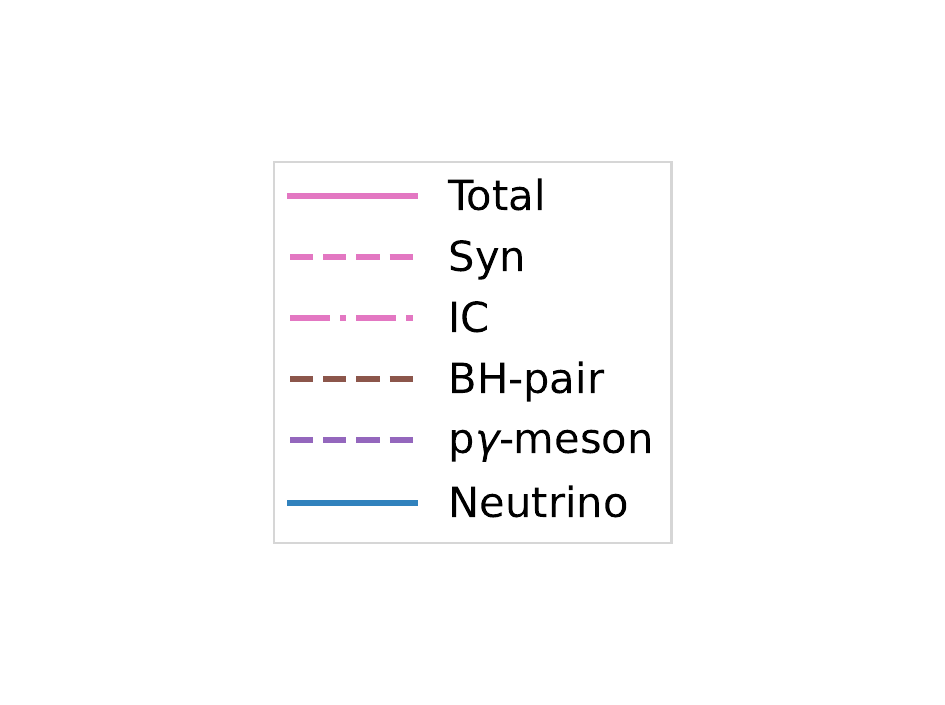}\\
\includegraphics[width=0.32\linewidth]{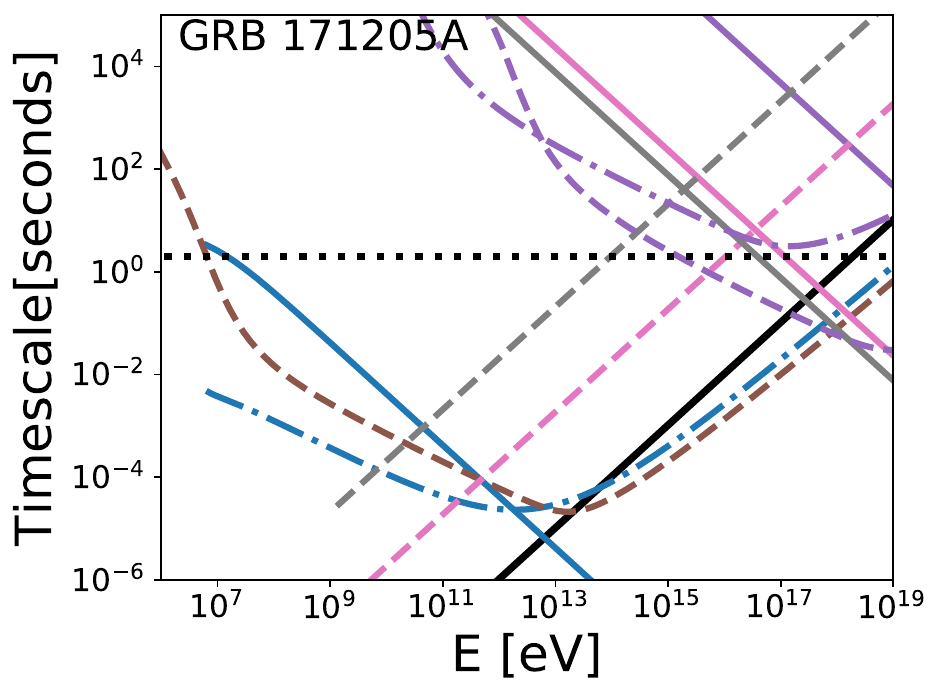}
\includegraphics[width=0.32\linewidth]{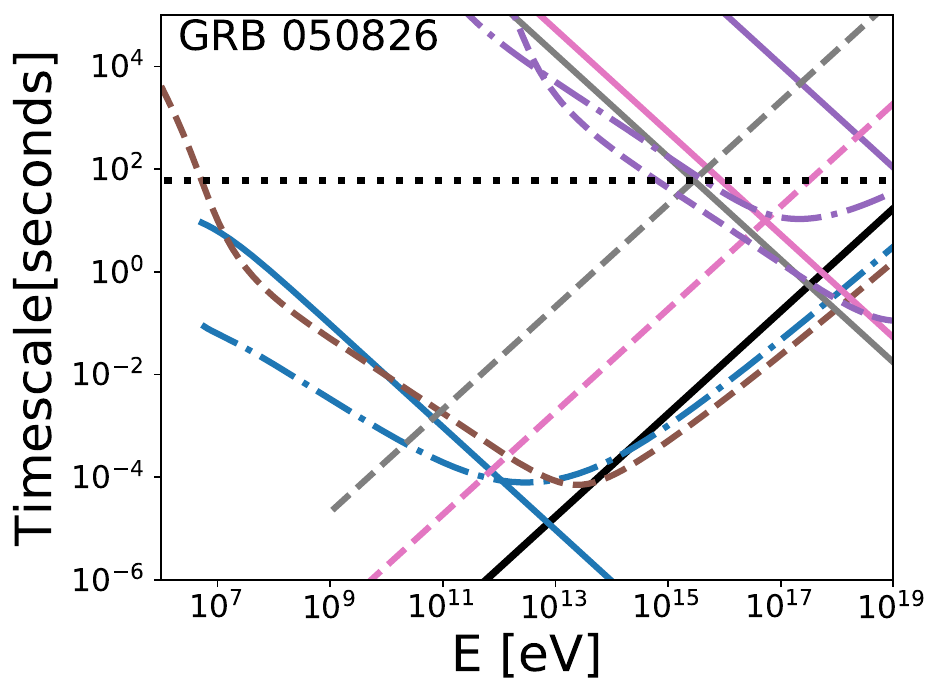}
\includegraphics[width=0.345\linewidth]{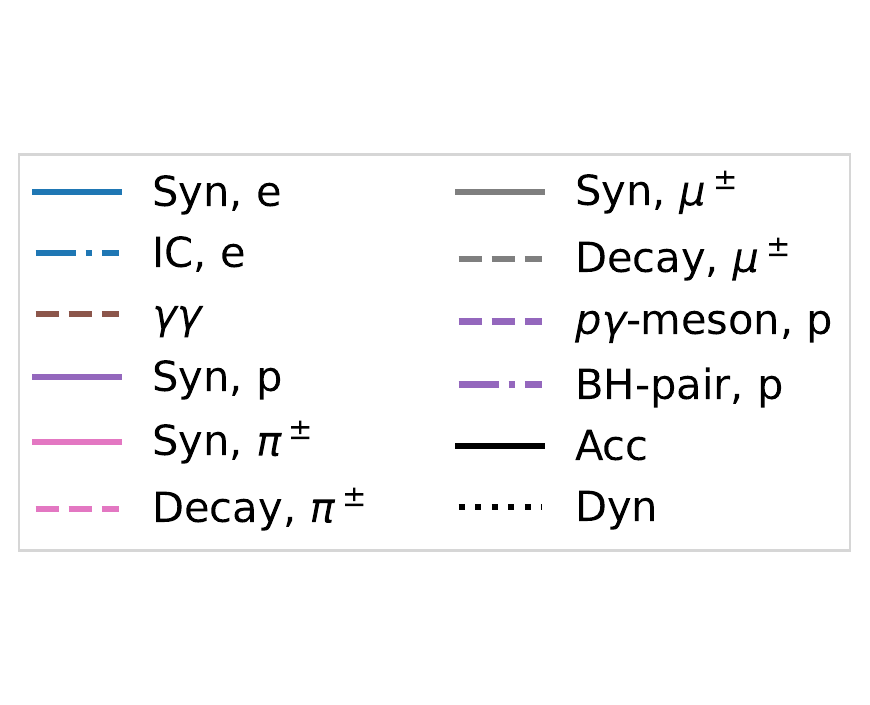}
\caption{Individual fits of GRB~171205A and GRB~050826. \fermilat{} upper limits are shown as red arrows, and \swiftbat{} data are shown in black. The total predicted photon spectra at the posterior median parameter values are shown as solid pink lines with 1\,$\sigma$ uncertainty bands, and contributions from hadronic processes are indicated by dashed purple and brown lines. Model predicted neutrino spectra at the posterior median parameter values are shown as solid blue lines, with shaded regions representing 1\,$\sigma$ uncertainties. Timescales are calculated using posterior median parameter values listed in Table~\ref{tab:best_fit_params}.}
    \label{fig:17_05}
\end{figure*}

\begin{figure*}[!tbh]
\centering
\includegraphics[width=0.32\linewidth]{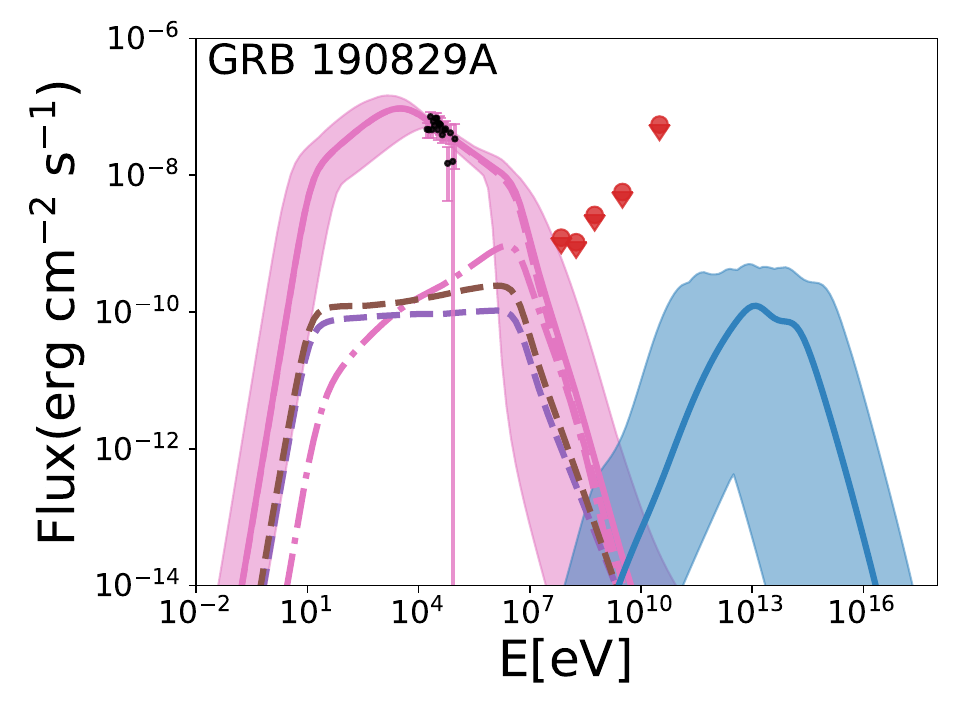}
\includegraphics[width=0.32\linewidth]{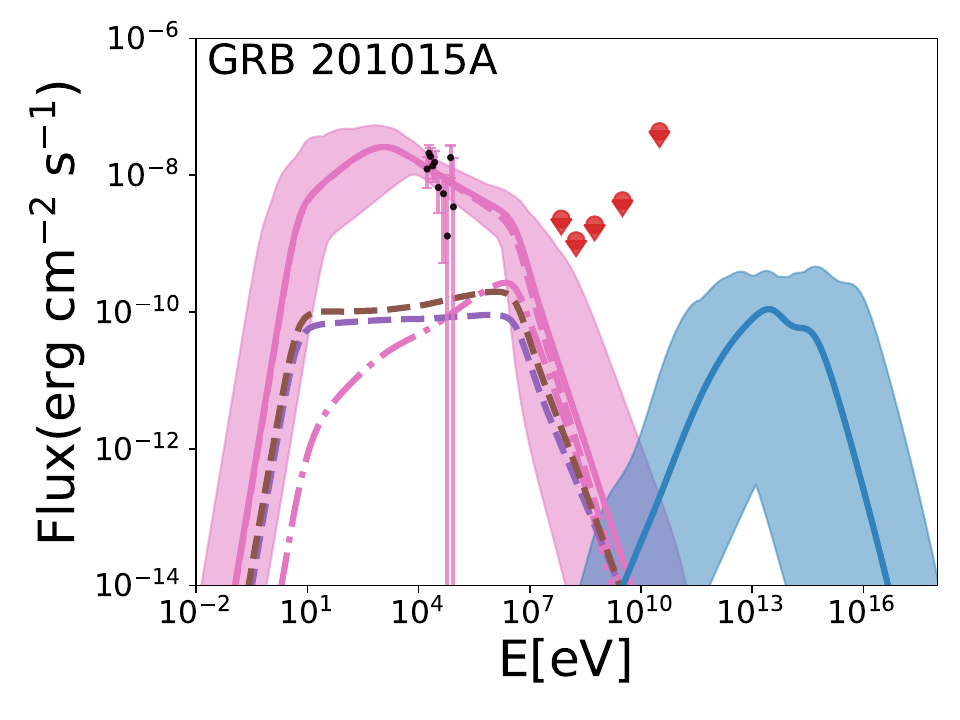}
\includegraphics[width=0.32\linewidth,trim=2cm 1cm 2cm 2cm,clip]{legend_only.pdf}\\
\includegraphics[width=0.32\linewidth]{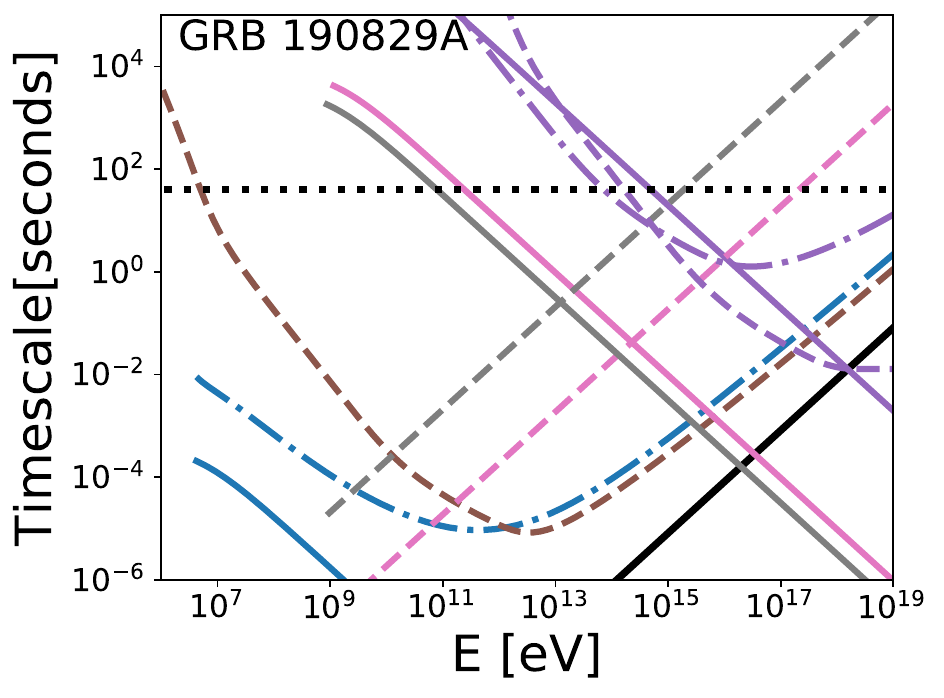}
\includegraphics[width=0.32\linewidth]{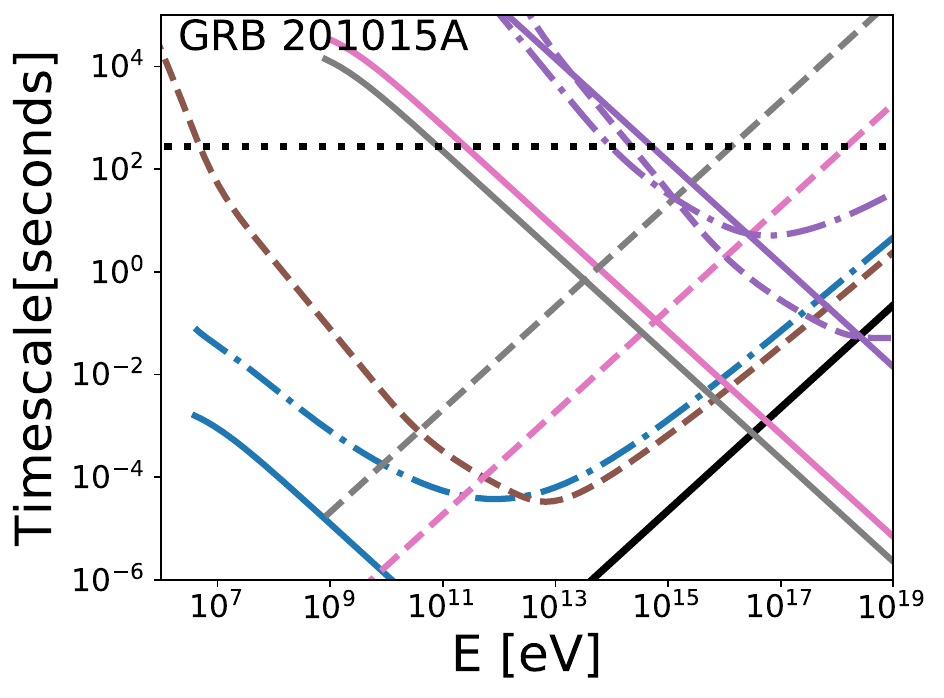}
\includegraphics[width=0.345\linewidth]{timescale_legend_only.pdf}
\caption{Same as Fig.~\ref{fig:17_05}, but for GRB~190829A and GRB~201015A. The predicted photon and neutrino spectra, as well as physical timescales, are calculated using the posterior median parameter values listed in Table~\ref{tab:best_fit_params}.}
\label{fig:19_20}
\end{figure*}

\begin{figure*}
\centering
\includegraphics[width=0.9\linewidth,trim=0cm 4cm 0.2cm 0cm,clip]{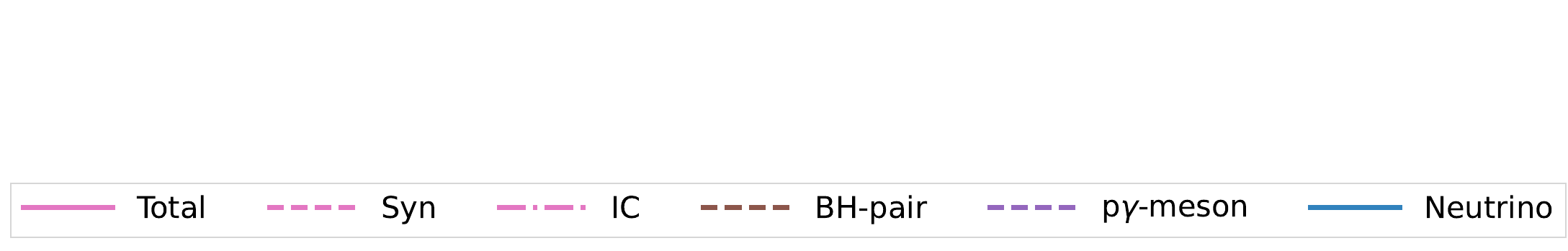}
\includegraphics[width=0.32\linewidth]{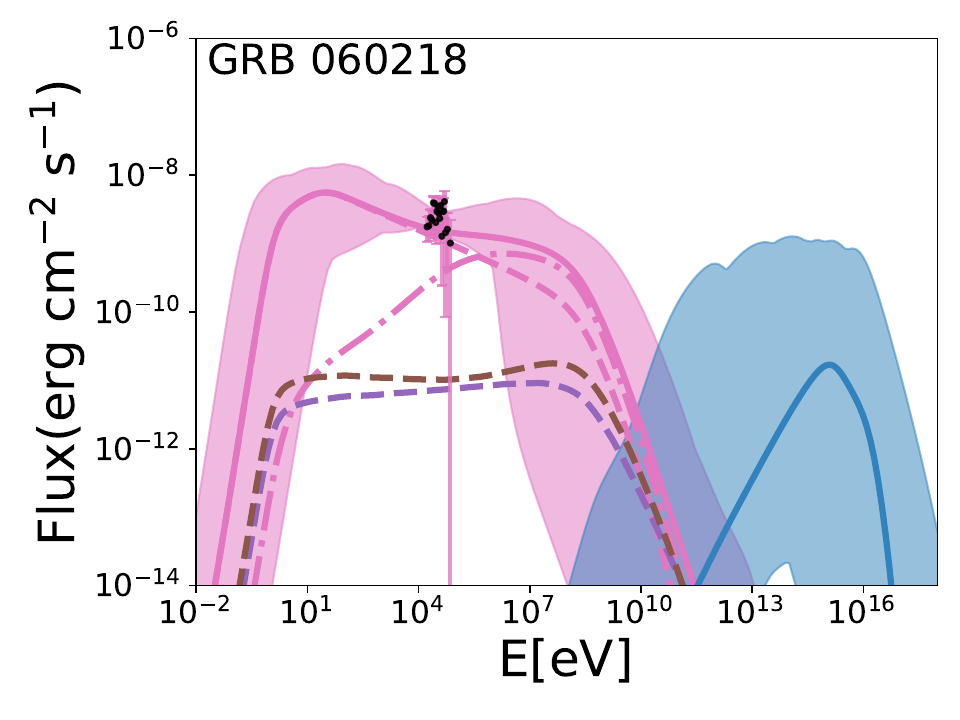}
\includegraphics[width=0.32\linewidth]{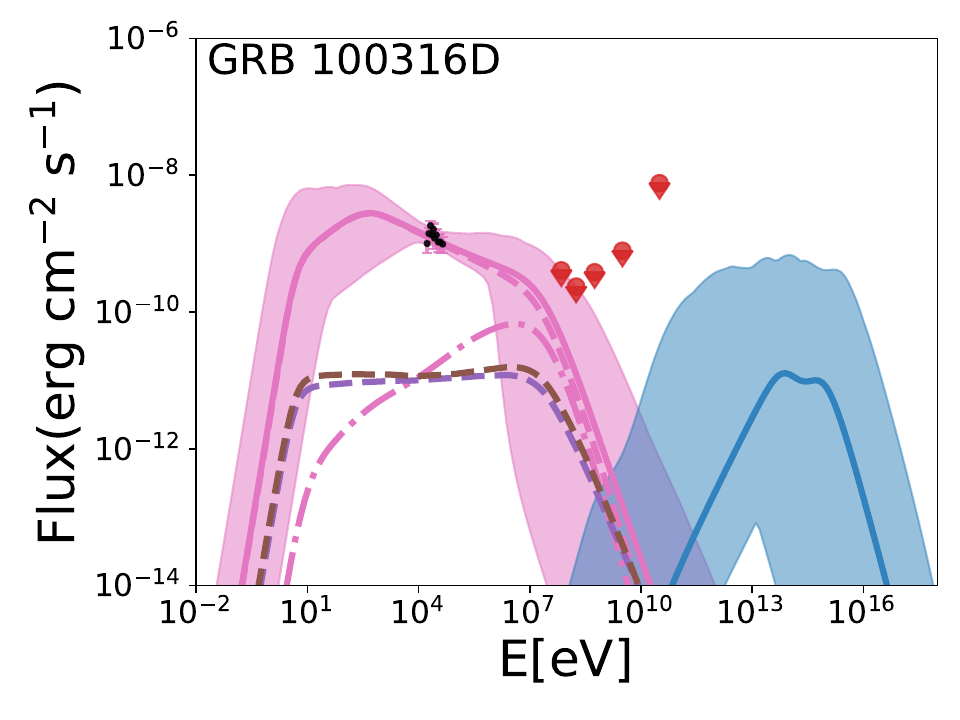}
\includegraphics[width=0.32\linewidth]{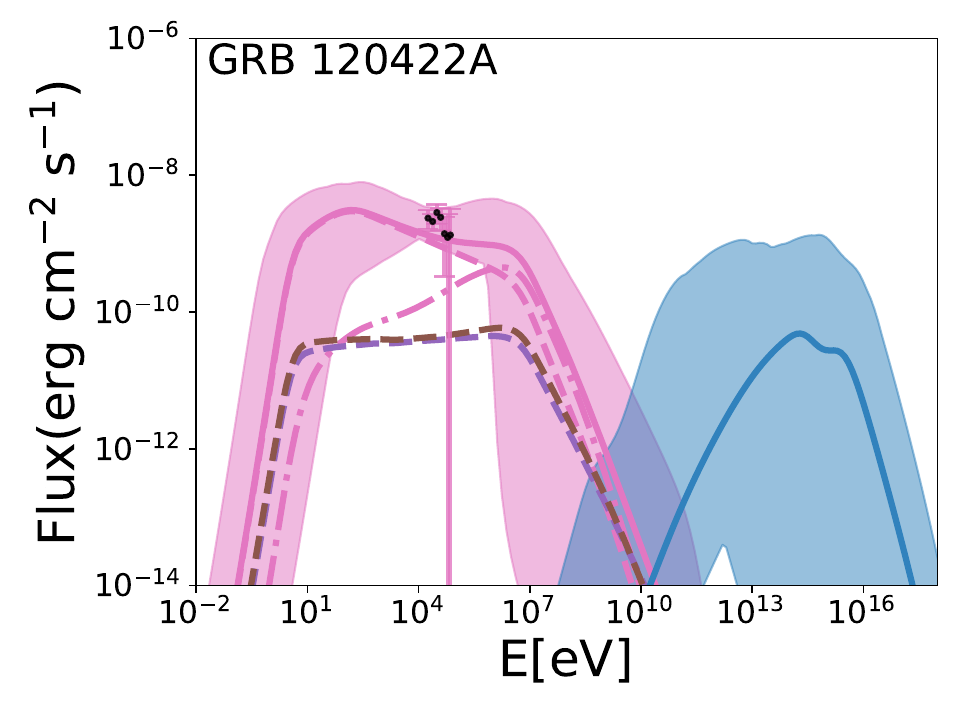} \\
\includegraphics[width=0.9\linewidth]{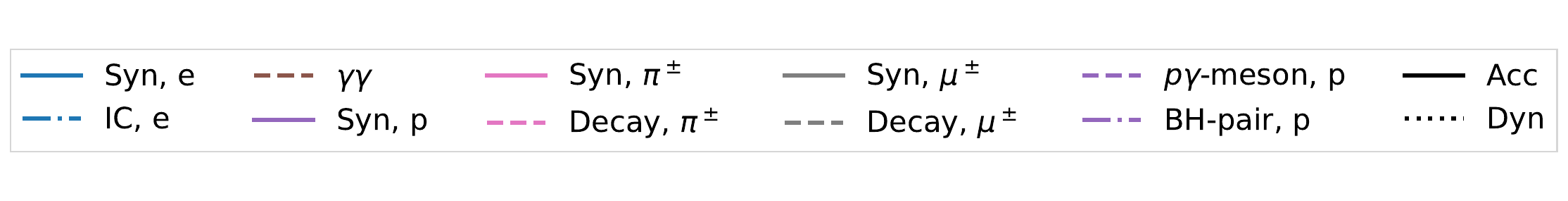}
\includegraphics[width=0.32\linewidth]{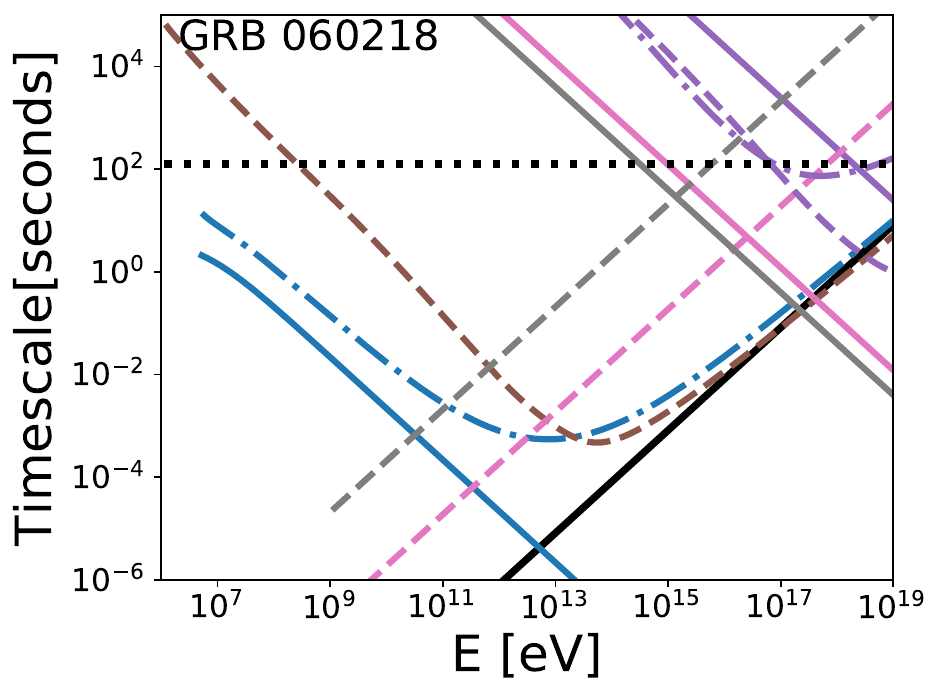}
\includegraphics[width=0.32\linewidth]{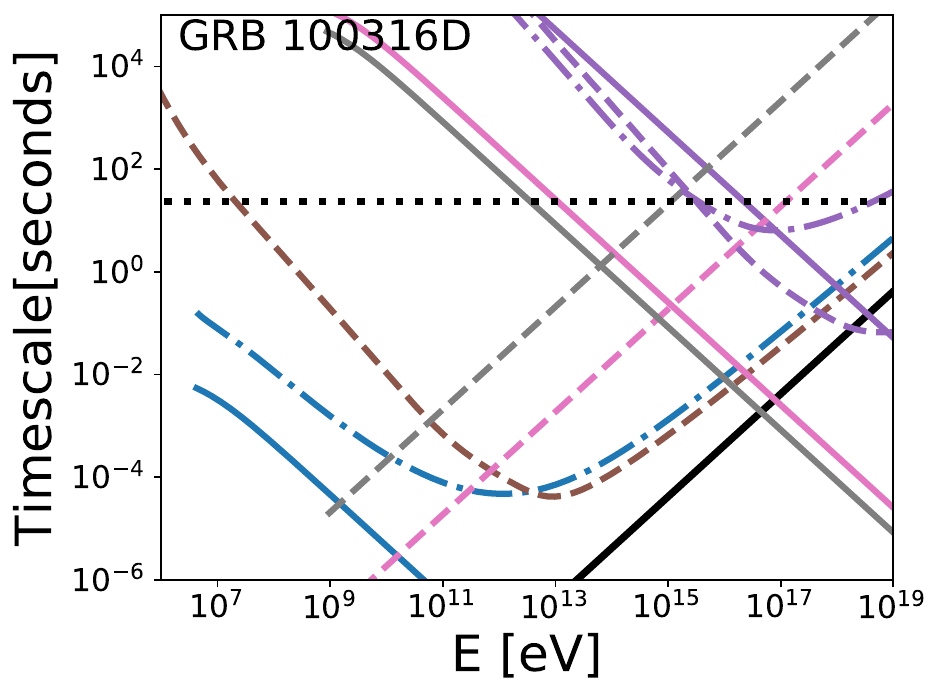}
\includegraphics[width=0.32\linewidth]{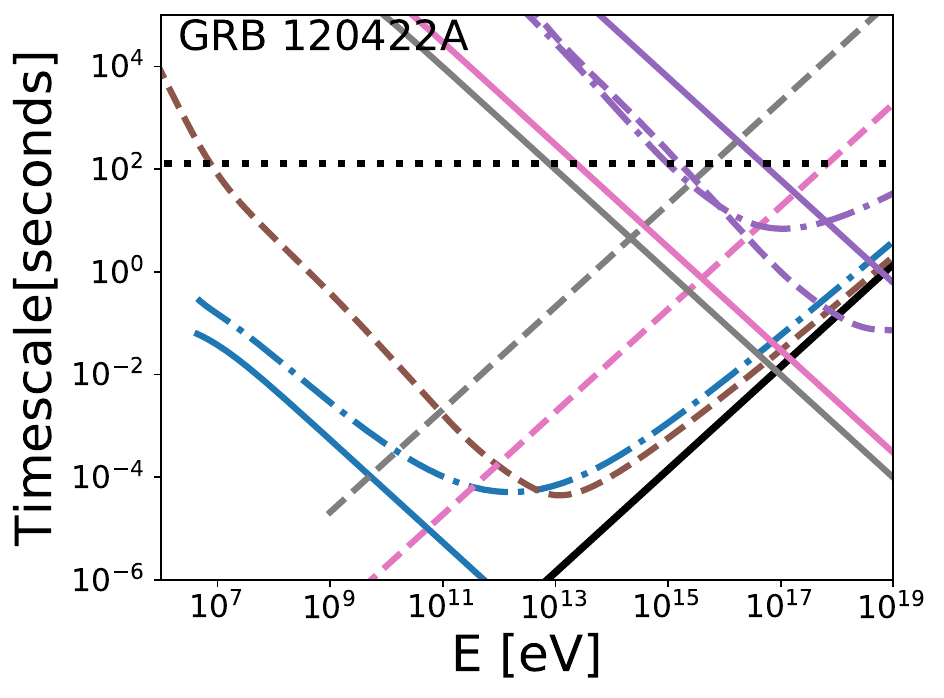}\\
\caption{Same as Fig.~\ref{fig:17_05}, but for GRB~060218, GRB~100316D, and GRB~120422A.}
\label{fig:06_10_12}
\end{figure*}

Since our models focus on the prompt emission, all discussions and comparisons with previous studies are likewise restricted to this interval. Because the parameter inference is performed within a Bayesian framework, for each parameter we quote the posterior median together with its 68\% credible interval, derived from the 1$\sigma$ range of the marginalized posterior distribution. These values characterize the central region of the posterior and serve as representative points within the parameter space explored by the MCMC sampler. The resulting parameter constraints are summarized in Table~\ref{tab:best_fit_params} and are used to generate the predicted neutrino and $\gamma$-ray spectra shown in Fig.~\ref{fig:17_05}--\ref{fig:06_10_12}. The contour plots of the MCMC samplers, with the 1D marginalized posterior probability distributions of the seven free parameters after chain convergence and discarding the burn-in phase, can be found in Figs.~\ref{fig:contour-05-17}--\ref{fig:contour-12} in Appendix~\ref{appendix:contour}. 

The predicted spectra show good agreement with the X-ray data and remain below the $\gamma$-ray upper limits. 
The decompositions of the total predicted photon spectra are shown for each LL~GRB at their posterior median parameter values. As discussed in Sec.~\ref {sec:model}, photon production involves various dynamical processes dominating at different energies, and the decompositions help reveal the intrinsic differences among the LL~GRBs. We show the primary contributions to the total photon spectra from leptonic processes, such as synchrotron emission and IC scattering from accelerated electrons, and from leptohadronic processes, including BH-pair cascades and $p\gamma$-meson interactions. The timescales of these physical processes are plotted alongside the predicted spectra in Figs.~\ref{fig:17_05}--\ref{fig:06_10_12}, illustrating how different mechanisms regulate photon and neutrino production.

Even though the absence of high-energy $\gamma$-ray and neutrino detections provides no direct constraint on leptohadronic channels, the \swiftbat{} X-ray data and $\gamma$-ray upper limits place strong constraints on both the total energy budget and the spectral shape of photon emission, which are tightly linked to leptohadronic processes. In particular, these data constrain how the energy of accelerated primary and secondary electrons is partitioned among synchrotron emission, IC scattering, and $\gamma\gamma$ absorption.

The hard X-ray spectra of GRB~171205A and GRB~050826 indicate a substantial IC contribution, as illustrated in Fig.~\ref{fig:17_05}. The IC cooling timescale is shorter than the synchrotron timescale in the Thomson regime, corresponding to a higher cooling rate via IC scattering. This behavior favors a relatively weaker magnetic field (i.e., a lower $\xi_B/\xi_e$) in these LL~GRBs (see Fig.~\ref{fig:param_xi_peB}) and results in $\xi_B$ being unconstrainedly small (see Appendix~\ref{fig:contour-05-17} for details).

A dominant IC component also requires a dense seed-photon field, implying a compact emission region with a small $R$. These same seed photons serve as targets for accelerated protons, enhancing both $p\gamma$-meson and BH-pair processes, which increases the expected neutrino flux and the high-energy photon output.
The connection between IC scattering and leptohadronic processes via target photons provides a strong constraint on the CR loading in the case of GRB~050218, even in the absence of $\gamma$-ray upper limits in the fit. Given its relatively high luminosity within our sample and the fact that its photon spectrum is well reproduced by IC-dominated emission, the results suggest that only a minor hadronic contribution is required, indicating that GRB~050218 may not represent a typical LL~GRB. Notably, although GRB~171205A is observationally low-luminosity, recent studies suggest that its physical origin may differ from that of typical LL~GRBs~\citep{2024ApJ...962..117L}. 
A contrasting behavior is seen in GRB~190829A and GRB~201015A, where synchrotron emission from accelerated electrons dominates, particularly at softer energies. The pronounced synchrotron component implies a strong magnetic field that efficiently cools charged particles, including electrons and CR protons, thereby limiting their maximum acceleration energy. It also strongly cools secondary pions and muons, whose decay times become longer at higher energies, as shown by the timescales in Fig.~\ref{fig:19_20}. 

In cases where the energy budget is more evenly partitioned, synchrotron and IC emissions reach comparable levels, as seen in GRB~060218, GRB~100316D, and GRB~120422A (Fig.~\ref{fig:06_10_12}). The similar values of $\xi_e$, $\xi_B$, and $\xi_p$ (see Fig.~\ref{fig:param_xi_peB}) suggest that the shock energy is nearly evenly partitioned among magnetic, electron, and proton energy densities, approaching the maximum acceleration efficiency expected for non-thermal particles in internal shocks~\citep{Kumar:2014upa, Sironi:2015oza, 2019MNRAS.485.5105C}.

For GRB~100316D, BH-pair and photomeson-induced cascade emission contribute significantly to the total photon flux. This arises from an intermediate magnetic energy density -- less extreme than in GRB~190829A and GRB~201015A -- allowing efficient proton interactions while maintaining strong radiative cooling. 
While GRB~060218 exhibits a similar energy partition, the absence of $\gamma$-ray upper limits results in weaker constraints on cascade emission and CR loading compared to GRB~100316D (see Fig.~\ref{fig:06_10_12}).

\begin{figure}
    \centering
\includegraphics[width=\linewidth]{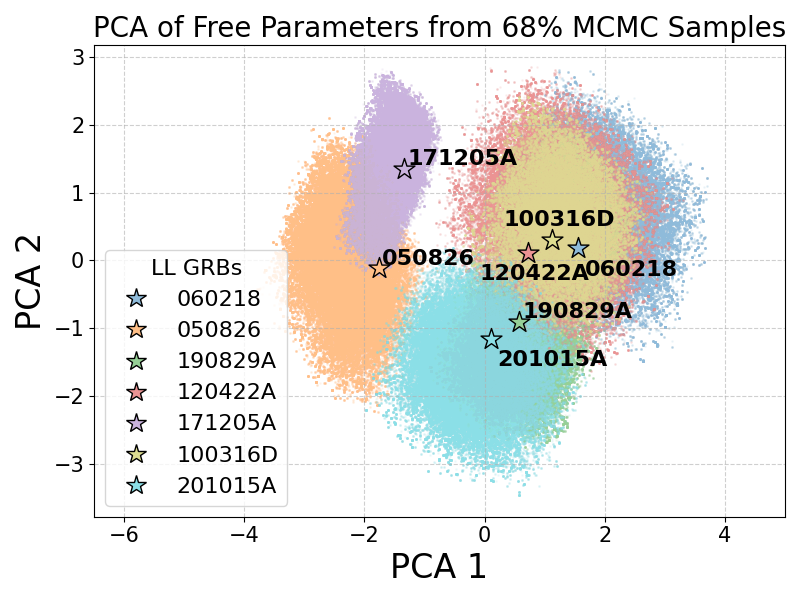}
    \caption{PCA projection onto the first two principal components of the seven LL~GRBs based on the 68\% confidence region of their post-burn-in MCMC samples for seven fitted parameters. Each colored cloud represents the 1$\sigma$ distribution of a GRB's fitted parameters, with individual points corresponding to sampled parameter sets and the posterior median parameter values marked by black circled spots.}
    \label{fig:PCA}
\end{figure}

To further quantify and visualize the similarities and differences among the seven LL~GRBs, we performed Principal Component Analysis (PCA), a machine-learning-based dimensionality-reduction technique, on the 1$\sigma$ subset of the post-burn-in MCMC samples of their seven fitted parameters to capture the primary variations while accounting for parameter uncertainties. PCA transforms the correlated variables into a smaller set of uncorrelated components that represent the primary directions of variance in the parameter space. By projecting the MCMC samples onto the first two principal components (PCA\,1 and PCA\,2), we obtain a compact two-dimensional representation for each LL~GRB. The resulting PCA plot (Fig.~\ref{fig:PCA}) shows clustered distributions that are broadly consistent with the physical interpretations discussed above: LL~GRBs with strong synchrotron components (GRB~190829A and GRB~201015A) occupy regions distinct from those dominated by IC (GRB~050826 and GRB~171205A) or cascade emission. Each point within a cloud corresponds to a single parameter set from the MCMC posterior, while the best-fit parameter sets are marked by black circled points. This visualization highlights both the similarities and the intrinsic diversity among the seven LL~GRBs.
The details in terms of their energy partitioning and emission properties are discussed as follows.

\begin{figure}\centering\includegraphics[width=\linewidth]{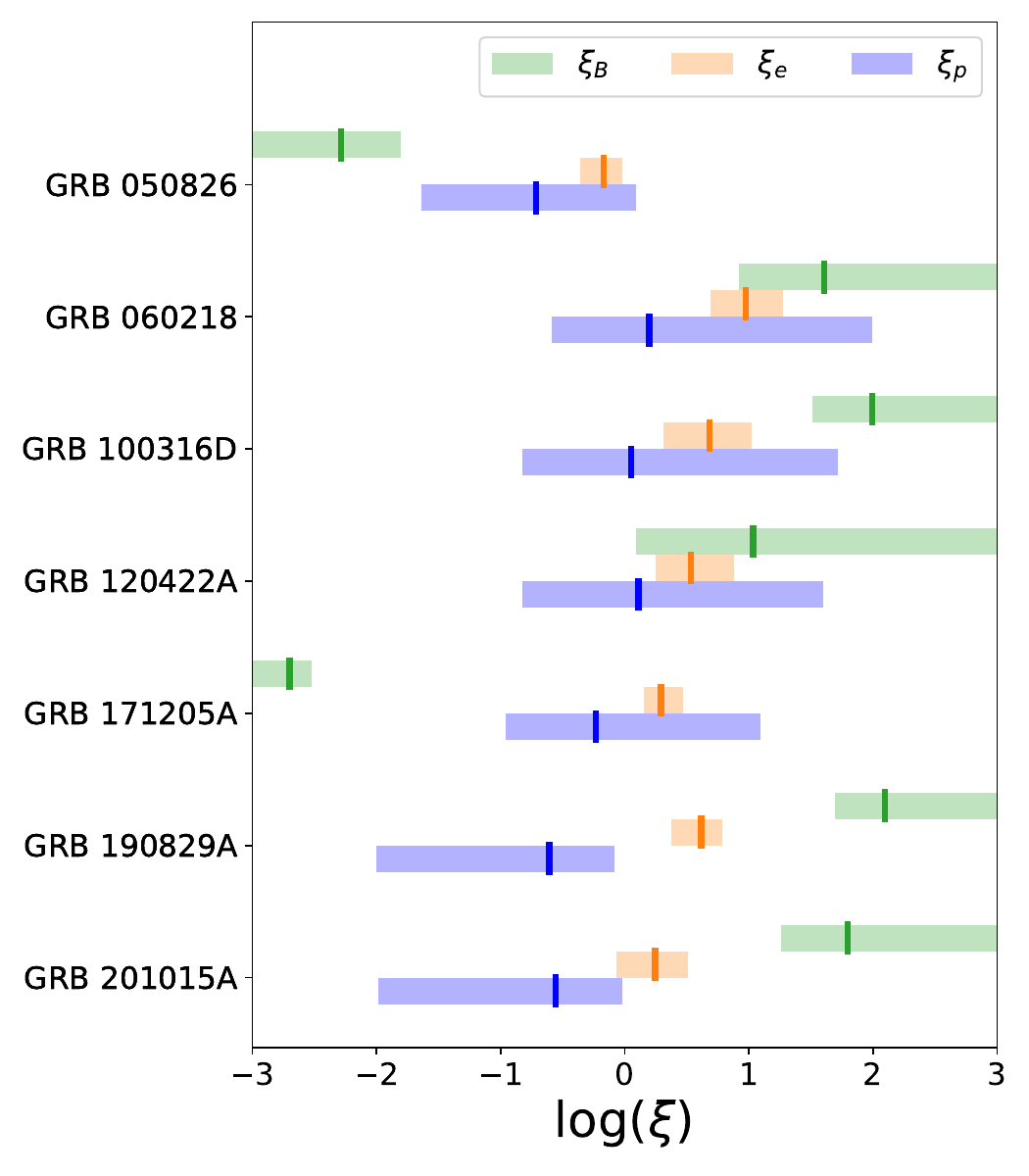}
    \caption{Visualization of the best-fit and $1\sigma$ uncertainties of $\xi_p$ (blue band), $\xi_e$ (yellow band) and $\xi_B$ (green band) for seven LL~GRBs.}
    \label{fig:param_xi_peB}
\end{figure}
The $1\,\sigma$ confidence intervals of the CR loading factor, $\xi_p$, span 
$\log(\xi_p) \sim -2.03$--$2.00$
across the sample, as summarized in Table~\ref{tab:best_fit_params} and illustrated in Figure~\ref{fig:param_xi_peB}.
The posterior median values of $\xi_p$ are found for the two events likely associated with the SBO scenario, GRB~060218 and GRB~100316D, which yield $\xi_p\sim1.3$-$1.6$.
The lower end of $\xi_p$ is constrained by GRB~190829A and GRB~201015A
, consistent with their synchrotron-dominated photon spectra and higher $\xi_B$, which suppress IC and cascade contributions as also constrained by the \fermilat{} upper limits.
GRB~050826 likewise exhibits a small $\xi_p$, but as discussed earlier, the IC-dominated spectrum is well reproduced by purely leptonic processes, implying that only a minor hadronic component is required.

\begin{figure}
    \centering    \includegraphics[width=\linewidth]{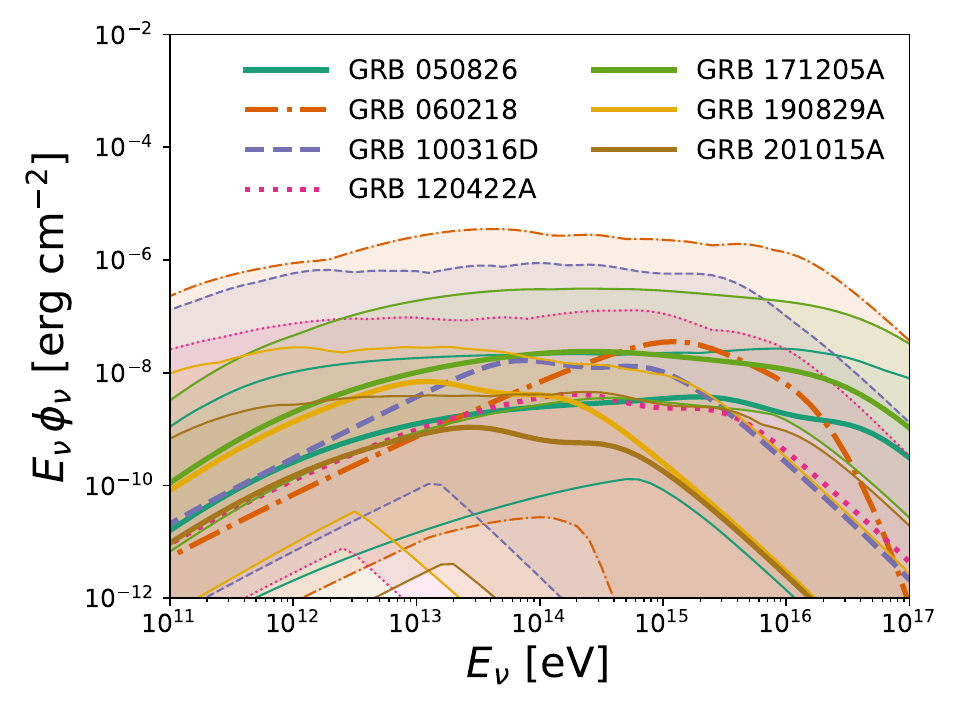}
    \caption{Predicted neutrino fluence (1$\sigma$ confidence region) for seven LL~GRBs, calculated using the posterior median parameter values listed in Table~\ref{tab:best_fit_params}.
    }
    \label{fig:neutrino-fluence}
\end{figure}
The predicted all-flavor neutrino fluences, integrated over the prompt emission durations $T_{90}$, are shown in Fig.~\ref{fig:neutrino-fluence}. GRB~060218 and GRB~100316D, characterized by the longest durations and relatively high CR loading factors, yield the highest expected fluences among the sample. 

\begin{table*}[!tbh]
\centering
\caption{Expected total number of muon and anti-muon neutrino events for LL~GRBs in current and future neutrino telescopes. Values are calculated for neutrinos with energies above 50~TeV and are given in units of $10^{-4}$. Each entry is given as the central value with the 16\%--84\% range in brackets, i.e., [lower bound, upper bound]. The effective areas for IceCube, IceCube-Gen2, HUNT, and TRIDENT are evaluated at the declination of each LL~GRB; for KM3NeT, P-ONE, and Baikal-GVD the best-scenario effective areas are used.}
\label{tab:nu_num}
\scriptsize;
    \begin{tabular}{lccccccc}
    \hline
    GRB & IceCube & IceCube-Gen2 & KM3NeT & P-ONE & HUNT & TRIDENT & GVD\\
    \hline
    050826 & $0.12\,[0.0,0.9]$ & $0.92\,[0.0,7.1]$ & $0.18\,[0.0,1.4]$ & $0.74\,[0.0,5.8]$ & $1.29\,[0.0,10.4]$ & $0.78\,[0.0,6.0]$ & 0.002\,[0.0,0.017]\\
    060218 & $0.71\,[0.0,116.2]$ & $4.37\,[0.0,823.3]$ & $1.04\,[0.0,164.1]$ & $4.01\,[0.0,739.7]$ & $5.56\,[0.0,1439.0]$ & $3.69\,[0.0,718.5]$  & 0.014\,[0.0,1.979]\\
    100316D & $0.03\,[0.0,2.0]$ & $0.33\,[0.0,19.8]$ & $0.77\,[0.0,46.0]$ & $3.41\,[0.0,199.0]$ & $7.13\,[0.0,402.9]$ & $3.11\,[0.0,185.3]$  & 0.009\,[0.0,0.533] \\
    120422A & $0.15\,[0.0,4.9]$ & $1.49\,[0.0,49.9]$ & $0.20\,[0.0,6.6]$ & $0.85\,[0.0,27.8]$ & $1.75\,[0.0,54.1]$ & $0.90\,[0.0,30.1]$   & 0.002\,[0.0,0.078]\\
    171205A & $0.49\,[0.1,6.8]$ & $0.58\,[0.1,8.1]$ & $1.35\,[0.2,18.4]$ & $5.69\,[0.8,78.8]$ & $11.00\,[1.5,150.0]$ & $5.36\,[0.7,73.5]$  & 0.016\,[0.0,0.223]\\
    190829A & $0.06\,[0.0,0.4]$ & $0.41\,[0.0,2.1]$ & $0.06\,[0.0,0.3]$ & $0.67\,[0.0,3.9]$ & $1.55\,[0.0,8.2]$ & $0.28\,[0.0,1.4]$  & 0.001\,[0.0,0.009]\\
    201015A & $0.01\,[0.0,0.1]$ & $0.05\,[0.0,0.3]$ & $0.01\,[0.0,0.1]$ & $0.14\,[0.0,1.0]$ & $0.31\,[0.0,2.1]$ & $0.12\,[0.0,0.9]$   & 0.0\,[0.0,0.002]\\
    \hline
    \end{tabular}%
\tablecomments{References: IceCube~\citep{2021arXiv210109836I}, IceCube-Gen2~\citep{IceCubeGen2TDR}, KM3NeT~\citep{2024EPJC...84..885K}, P-ONE~\citep{2020NatAs...4..913A}, HUNT~\citep{Huang:2025x5}, Trident~\citep{2022arXiv220704519Y}, Baikal-GVD~\citep{Malyshkin:2023qnc}.}
\end{table*}

Using the effective areas of current and next-generation neutrino detectors, we estimate the expected numbers of muon and anti-muon neutrino events above 50~TeV for each LL~GRB (see Table~\ref{tab:nu_num}). Even in the most optimistic scenario given by the upper bounds, the current facilities (IceCube and KM3NeT) are unlikely to detect individual LL~GRBs, with expected event numbers $\lesssim 0.01$ per burst. Next-generation observatories such as IceCube-Gen2, KM3NeT (after full commissioning), P-ONE, HUNT, and TRIDENT improve the sensitivity by roughly an order of magnitude, reaching $\sim 0.1$ events per nearby LL~GRB.

Although the individual searches are below the single-event threshold, stacking analyses leveraging larger samples, especially those identified by wide-field missions like Einstein Probe, could yield statistically significant signals. Equally important are model-based theoretical predictions, such as those presented in this work, which provide physically motivated templates for stacking analyses and multi-messenger correlation studies.
Such coordinated observational and theoretical efforts will be essential to test whether LL~GRBs contribute a subdominant yet distinct component to the diffuse astrophysical neutrino background.

\section{Summary and perspective}\label{sec:summary} 

In this study, we modeled the prompt emission of seven LL~GRBs by simultaneously evolving nonthermal electrons and protons while analytically updating the target photon field using {\sc AMES}. By fitting the observed X-ray and $\gamma$-ray data, we constrained key physical parameters, including the energy partition in accelerated protons ($\xi_p$) and electrons ($\xi_e$), and magnetic fields ($\xi_B$), along with other parameters describing the acceleration and dissipation. More importantly, the distinct preferred parameter ranges inferred for each LL~GRB indicate the presence of potential sub-populations governed by different dominant physical processes. This trend is further supported by our PCA analysis. Although LL~GRBs are uniformly characterized by their low luminosities, the underlying physical mechanisms responsible for their faint prompt emission can vary substantially from event to event. To identify LL~GRBs that are capable of efficiently accelerating cosmic rays and producing neutrinos, our results suggest that the most promising candidates are those that are intrinsically low-luminosity---rather than dimmed by off-axis effect---with relatively balanced energy partition, in particular a sufficiently large cosmic-ray loading factor, and with durations long enough to enable efficient neutrino production. GRB~060218 and GRB~100316D exemplify such events. The constrained model also enables predictions of the accompanying high-energy neutrino emission, providing physically motivated expectations for time-dependent stacking analyses in current and next-generation neutrino observatories.

In our sample, GRB~060218 and GRB~100316D, with $T_{90} \sim 2100~\mathrm{s}$ and $\sim 1300~\mathrm{s}$ respectively, have been primarily interpreted as SBO-driven events, while GRB~120422A, GRB~171205A, GRB~190829A, and GRB~201015A, with $T_{90} \sim 10$-$200~\mathrm{s}$, have been suggested to be consistent with emerging jets with relatively low Lorentz factors ($\Gamma \sim 10$). Our results are broadly consistent with previous studies, but derived from a novel modeling approach combining multi-wavelength data and PCA analysis. We further find that the posterior median values of the cosmic-ray loading factor fall in the range $\xi_p \sim 0.2$-$1.6$ based on our detailed leptohadronic treatment. Although the limited energy coverage of the available multi-wavelength data prevents tighter constraints on certain model parameters, this work represents a first step toward quantitatively characterizing LL~GRB properties and establishing physically grounded expectations for their associated high-energy neutrino emission. In a companion paper, we apply this methodology to estimate the contribution of the LL~GRB population to the diffuse astrophysical neutrino flux. Although this study focuses on the prompt emission, leptohadronic processes in the afterglow phase are expected to produce neutrinos at higher energies (100~PeV-EeV), which we plan to explore in future work.

From an observational perspective, EP is expected to improve detection efficiency and has already identified fast X-ray transients, a fraction of which may correspond to LL~GRBs~\citep{Yuan:2025cbh}. Space missions such as the Space-based multi-band astronomical Variable Objects Monitor~\citep{2015arXiv151203323C} and \swiftbat{} also provide complementary X-ray and $\gamma$-ray coverage, increasing the likelihood of capturing faint and sub-threshold LL~GRBs. Future MeV telescopes will further narrow the parameter space for prompt emission, given that \fermilat{} has proven invaluable for constraining hadronic processes in this study. 
Real-time triggers and rapid follow-up are essential for faint, abundant transients, and improved coordination among multi-wavelength telescopes will be crucial. Even non-detections and upper limits are valuable for constraining physical parameters. Future neutrino observatories, in particular, should incorporate model-based strategies to optimize the detection and interpretation of specific source classes and emission scenarios.

Finally, the modeling approach presented here is broadly applicable to other astrophysical objects with multi-messenger observations. By simultaneously evolving particle productions and interactions, and iteratively updating photon fields, this method can provide robust predictions for neutrino emission and guide multi-messenger searches across diverse classes of transients and steady astrophysical sources. 

\appendix
\section{Observational Data and X-ray Analysis}\label{appendix:x-ray}

\begin{table*}[bth]
\centering
\caption{Observation information for the LL~GRB events. Swift-BAT data are shown in the first set of columns; Fermi-LAT time ranges are given in Mission Elapsed Time (MET) [s].}
\begin{tabular}{lcccccc}
\hline\hline
GRB & \multicolumn{3}{c}{Swift-BAT} & \multicolumn{2}{c}{Fermi-LAT} \\
\cmidrule(lr){2-4} \cmidrule(lr){5-6}
    & Observation Date [UTC] & Exposure [s] & ObsID & T$_{min}$ [MET, s] & T$_{max}$ [MET, s] \\
\hline
050826   & 2007-04-24T18:50:55 & 29.6  & 00152113000 & --- & --- \\
060218   & 2006-02-18T03:39:34 & 80    & 00191157000 & --- & --- \\
100316D  & 2010-03-16T12:40:56 & 521.9 & 00416135000 & 290436280.4 & 290436802.4 \\
120422A  & 2012-04-28T10:50:37 & 61.4  & 00520658000 & --- & --- \\
171205A  & 2017-12-05T07:17:00 & 190.5 & 00794972000 & 534151143.4 & 534152243.4 \\
190829A  & 2019-08-29T19:53:04 & 58.2  & 00922968000 & 588801316  & 588801374 \\
201015A  & 2020-10-15T22:46:35 & 9.8   & 01000452000 & 624495057.8 & 624496057.8 \\
\hline
\end{tabular}
\label{tab:obsdata}
\end{table*}
Table~\ref{tab:obsdata} summarizes the observations from \swiftbat{} and \fermilat{} used in Sec.~\ref{sec:mm}. The results of the \swiftbat{} spectral analysis and isotropic luminosity estimation are proved in Table~\ref{tab:x-ray_analysis}. The spectra were fit using a cutoff power-law (CPL) model over the energy range 1\,keV to 10\,MeV. 
The resulting best-fit parameters, their 1$\sigma$ uncertainties, and the isotropic $\gamma$-ray luminosity estimates are summarized in Table~\ref{tab:x-ray_analysis}. Our fits are consistent with those reported in the batgrbcat\footnote{\url{https://swift.gsfc.nasa.gov/results/batgrbcat/}}~\citep{Lien_2016}, including cases where certain parameters remain unconstrained due to limited data quality or restricted energy coverage.

\begin{table*}[!tbh]
\centering
\caption{Best-fit parameters and their 1$\sigma$ confidence intervals for the seven LL~GRBs.} 
\begin{tabular}{lcccccc}
\toprule
GRB & $\alpha$  & ${E_{\rm cutoff}}$ & $z$ &$L_{\gamma, \rm iso}$ [ergs/s] & Function & $\chi^2_{\mathrm{d.o.f.}}$ \\
\midrule
050826 & $ 1.0\pm 0.75$ & $500.0\pm6342.58 $ &0.297 & $1.4\times 10^{49}$ &CPL & $ 62.57/58$ \\
060218 & $-0.6\pm1.60$ & $11.9\pm7.02$ &0.033 & $1.2\times 10^{46}$ &CPL & $21.43/34$ \\
100316D & $1.0\pm1.79$ & $21.6\pm30.94$ & $0.06$ & $3.0\times10^{46}$ & CPL & $16.75/16$ \\
120422A & $1.2\pm2.11$ & $33.3\pm72.16$ & 0.283 & $1.8\times10^{48}$ & CPL & $23.48/29$ \\
171205A & $1.0\pm0.38$ & $134.1\pm119.72$ & 0.0368 & $8.3\times10^{46}$ & CPL&$62.94/59$\\
190829A & $1.0\pm 0.75$ & $25.2\pm13.3$ & 0.0785 &$2.2\times10^{48}$ & CPL & $29.25/39$ \\
201015A & $1.9\pm3.00$ & $22.7\pm49.11$ & $0.426$ & $5.9\times10^{49}$ &  CPL& $28.24/39$ \\
\bottomrule
\end{tabular}
\tablecomments{Best-fit and 1$\sigma$ interval of spectral index ($\alpha$), cutoff energy ($E_{\rm cutoff}$), and redshift ($z$) for each source, assuming a cutoff power-law (CPL) model. The isotropic $\gamma$-ray luminosity ($L_{\gamma, \mathrm{iso}}$) is estimated over the 1\,keV-10\,MeV energy range after correcting for redshift.  
The goodness of fit is given by $\chi^2_{\mathrm{d.o.f.}}$}
\label{tab:x-ray_analysis}
\end{table*}

\section{Appendix B}\label{appendix:contour}
We show the confidence contours of the spectral fitting parameters for the seven LL~GRBs to illustrate the posterior median parameter values and uncertainties. 

\begin{figure*}
    \centering\includegraphics[width=0.45\textwidth]{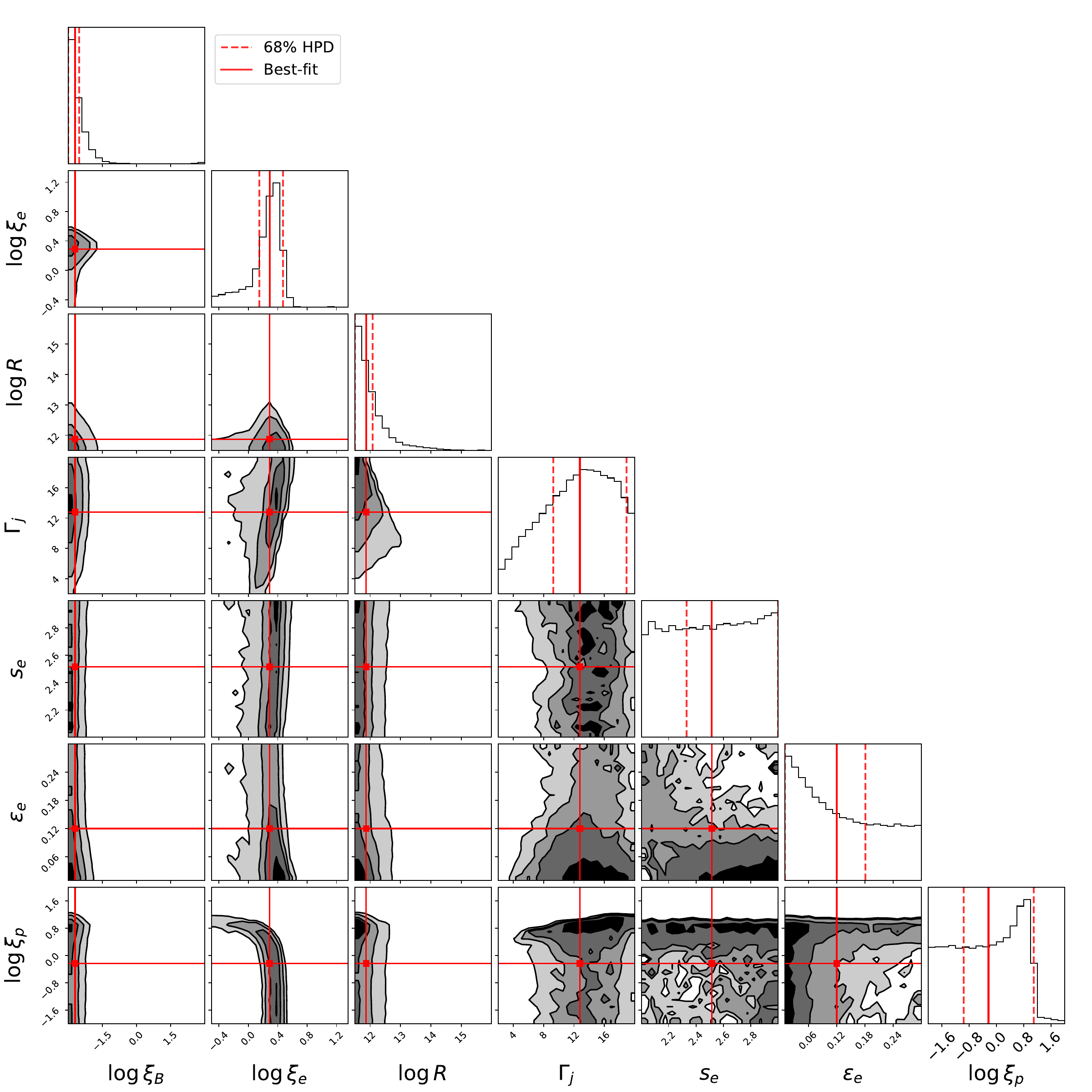}
        \includegraphics[width=0.45\textwidth]{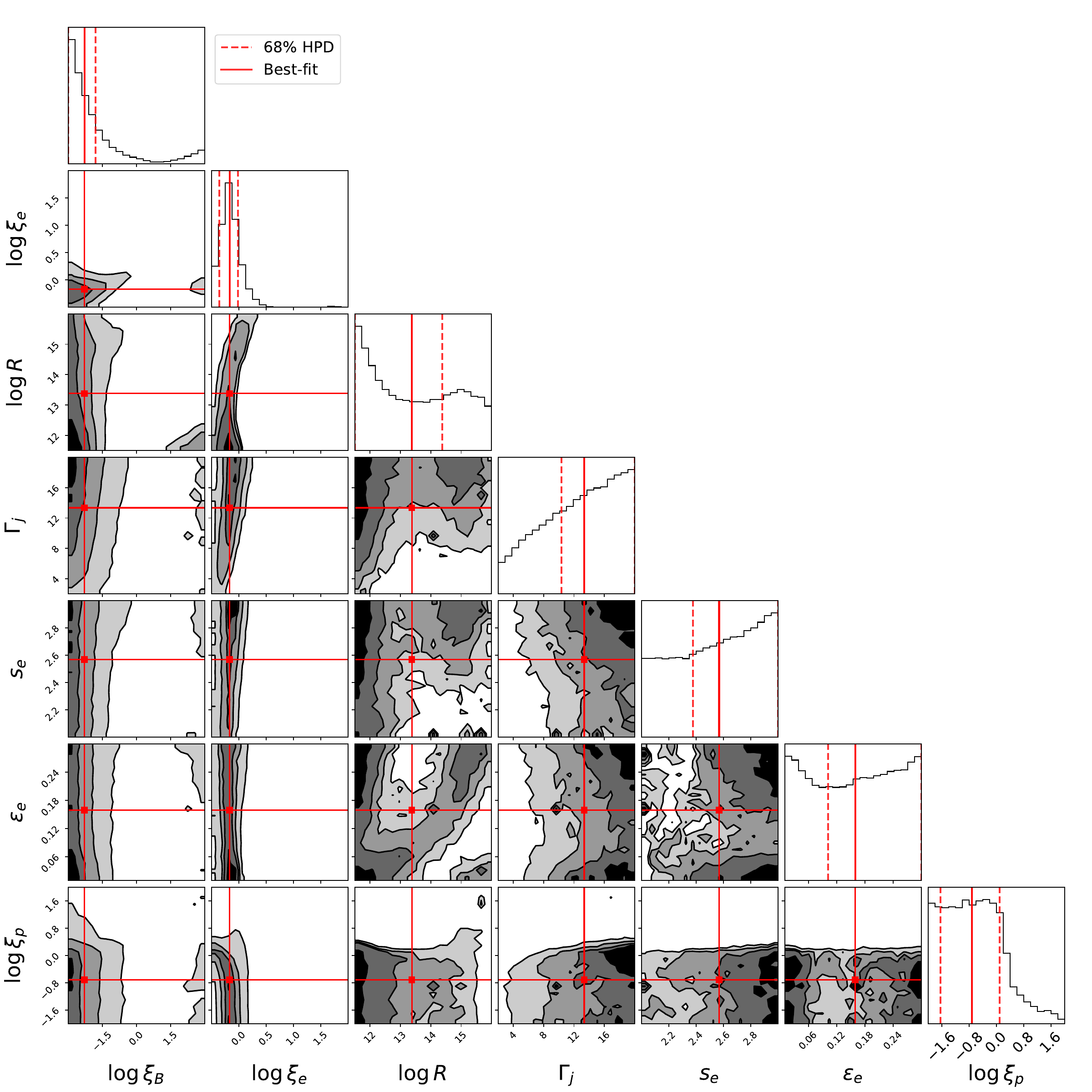}
    \caption{Confidence contours of the spectral fitting parameters for GRB~171205A (left) and GRB~050826 (right). The red points indicate the medians of the post-burn-in MCMC samples, while the red dashed lines mark the 1$\sigma$ highest posterior density (HPD) interval of the marginalized 1D posterior distribution. Fitted-parameter values are listed in Table~\ref{tab:best_fit_params}. }
    \label{fig:contour-05-17}
\end{figure*}

\begin{figure*}
    \centering
    
    \includegraphics[width=0.45\textwidth]{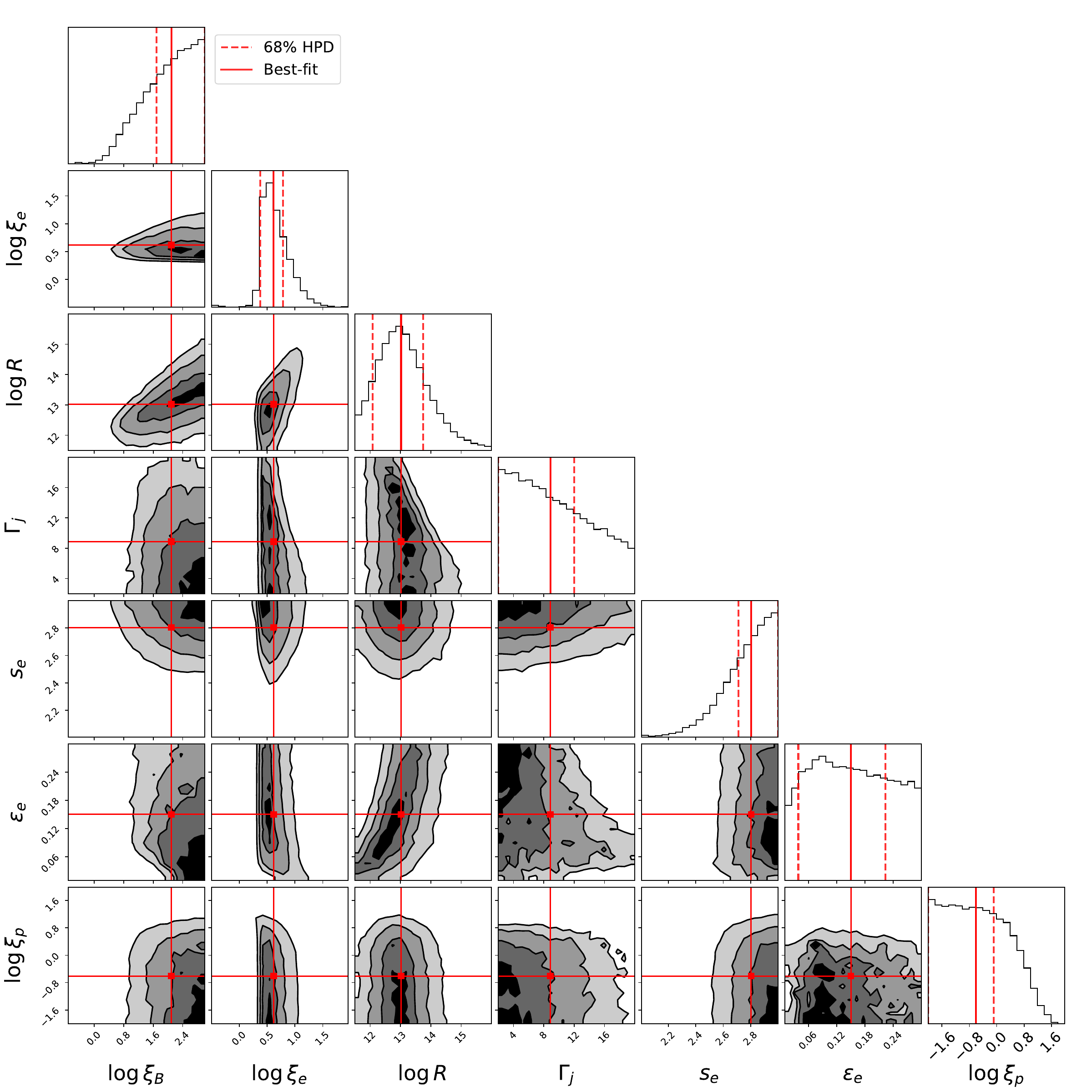}
    \includegraphics[width=0.45\textwidth]{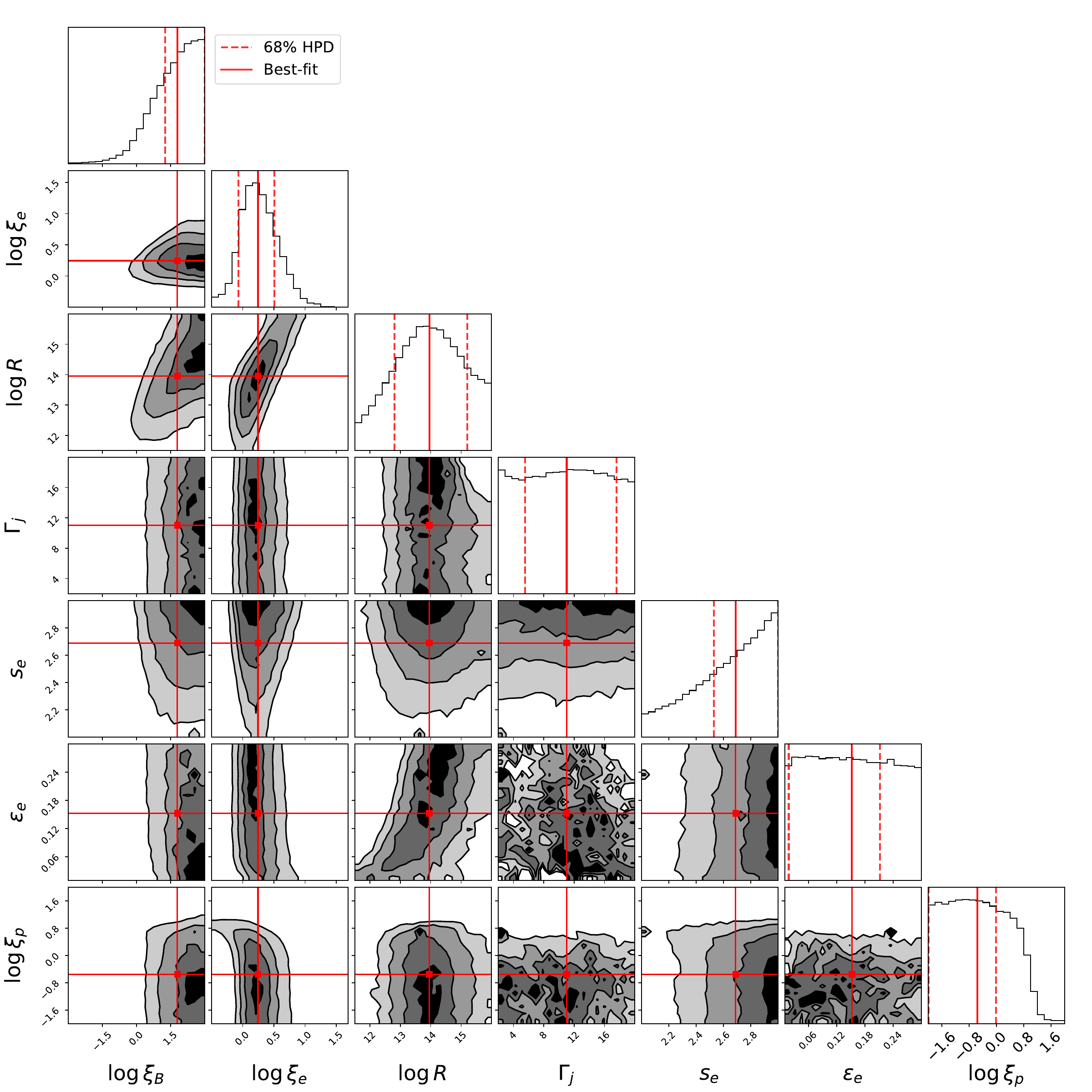}
    \caption{Same as Figure~\ref{fig:contour-05-17}, but for GRB~190829A (left) and GRB~201015A (right).}
    \label{fig:contour-19-20}
\end{figure*}

\begin{figure*}
    \centering
    \includegraphics[width=0.45\textwidth]{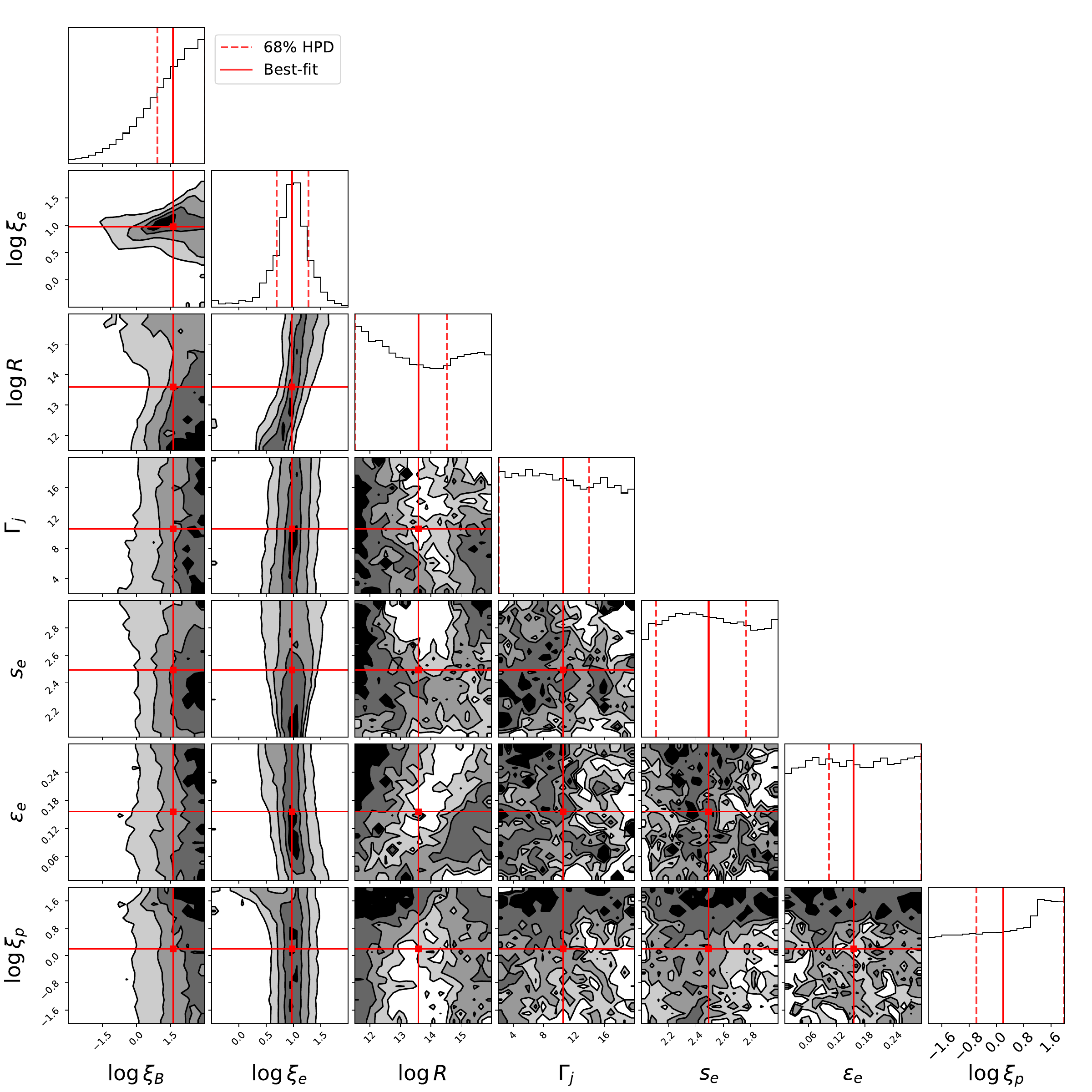}    \includegraphics[width=0.45\textwidth]{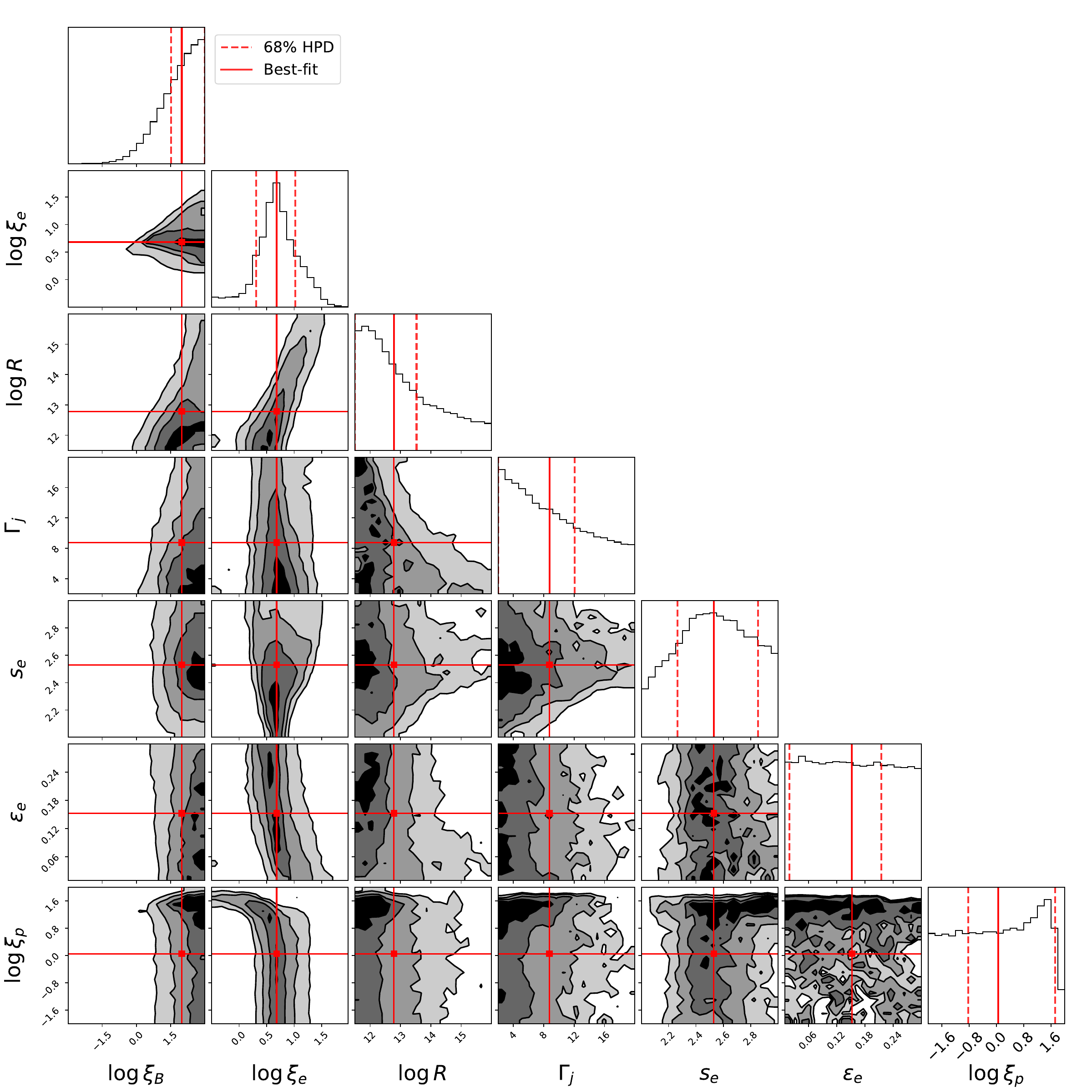}
    \caption{Same as Figure~\ref{fig:contour-05-17}, but for GRB~060218 (left) and GRB~100316D (right).}
    \label{fig:contour-06-10}
\end{figure*}

\begin{figure*}
    \centering
    \includegraphics[width=0.45\textwidth]{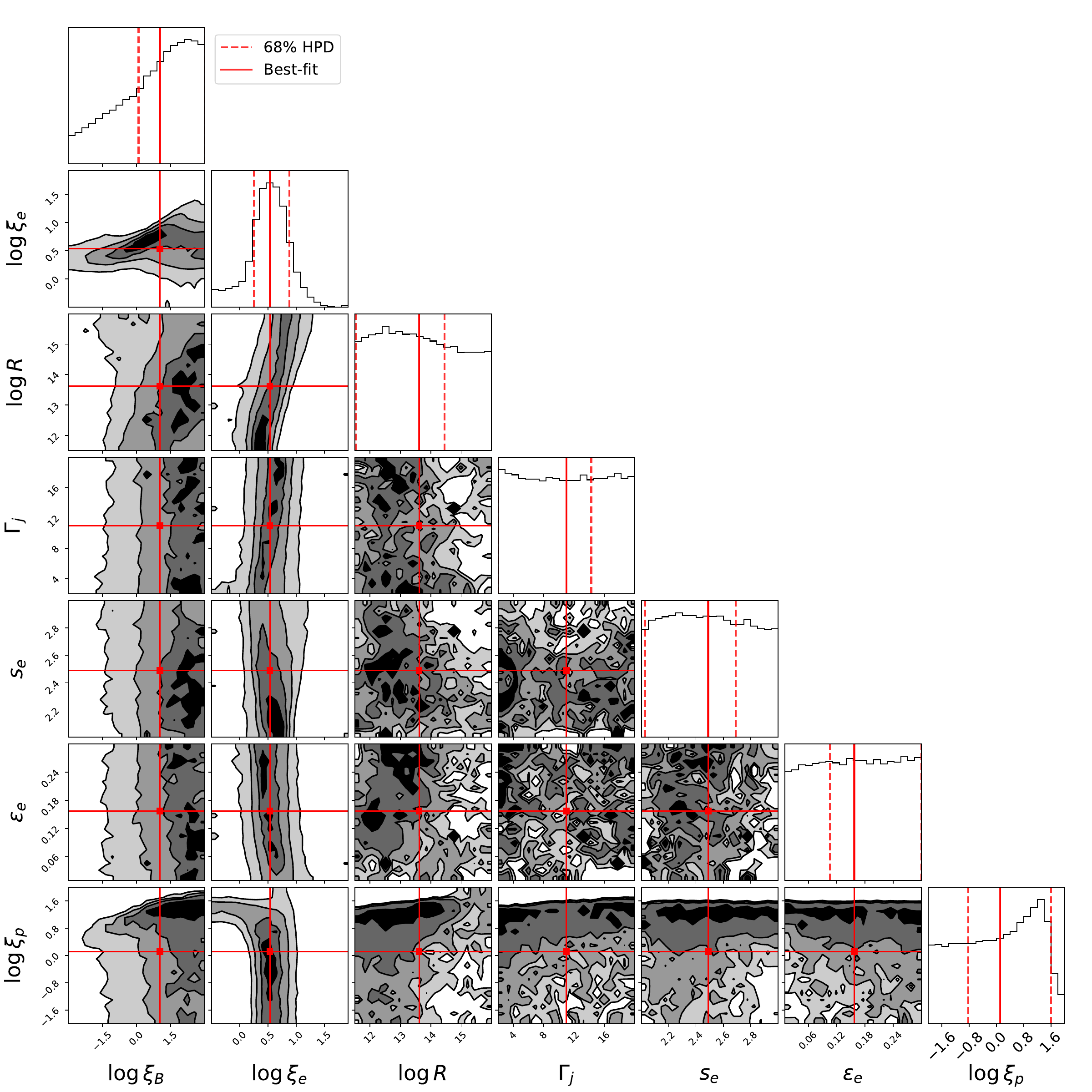}

    \caption{Same as Figure~\ref{fig:contour-05-17}, but for GRB 120422A.}
    \label{fig:contour-12}
\end{figure*}

\clearpage
\section{Acknowledgment}
We thank Kohta Murase, Carsten Rott, Xiangyu Wang, Chris Weaver, Shigeru Yoshida, Xuehui Yu, and Lihua Zhou for valuable scientific discussions and suggestions. We are grateful to the Baikal-GVD, HUNT, and TRIDENT collaborations, particularly Dmitry Zaborov; Mingjun Chen and Tianqi Huang; and Donglian Xu and Fuyudi Zhang, for providing detector effective areas. We also acknowledge the IceCube-Gen2 and KM3NeT collaborations for making their effective areas publicly available, and we thank Albrecht Karle and Jakob van Santen for guidance on the IceCube-Gen2 effective areas. B.T.Z. is supported in China by the National Key R\&D Program of China under the grant 2024YFA1611402. The support and resources from the Center for High Performance Computing at the University of Utah are gratefully acknowledged. This work benefited from publication support provided by the American Astronomical Society.

\bibliography{main}{}
\bibliographystyle{aasjournalv7}

\end{document}